\documentclass[12pt,english]{article}
\usepackage[T1]{fontenc}
\usepackage[latin9]{inputenc}
\usepackage{amsmath}
\usepackage{bbold}
\usepackage{graphicx}
\usepackage[algoruled,boxed,lined]{algorithm2e}
\usepackage[top=0.8in, bottom=0.8in, left=1in, right=1in]{geometry}

\usepackage{capt-of}

\usepackage{natbib}
\usepackage{babel}
\makeatletter
\g@addto@macro{\@algocf@init}{\SetKwInOut{Parameter}{Parameters}}
\makeatother

\usepackage{enumerate}

\usepackage{listings}
\usepackage{color} 
\definecolor{mygreen}{RGB}{28,172,0} 
\definecolor{mylilas}{RGB}{170,55,241}

\usepackage[section]{placeins}
\usepackage{authblk}

\usepackage{fancyhdr}
\fancypagestyle{alim}{\fancyhf{}\renewcommand{\headrulewidth}{0pt}\fancyfoot[R]{ {\footnotesize Correspondence to \textit{p.maybank@pgr.reading.ac.uk, i.bojak@reading.ac.uk, r.g.everitt@reading.ac.uk}}} }
\usepackage{floatpag}
\usepackage{xr}
\begin{document}


\title{Fast approximate Bayesian inference for stable differential equation models}

\author[1]{Philip Maybank}
\author[2]{Ingo Bojak}
\author[1]{Richard G. Everitt}
\affil[1]{Department of Mathematics \& Statistics, University of Reading}
\affil[2]{School of Psychology and Clinical Language Sciences, University of Reading}
\setcounter{Maxaffil}{0}
\renewcommand\Affilfont{\itshape\small}
\maketitle
\thispagestyle{alim}
\renewcommand{\headrulewidth}{0pt}
\fancyhead{}

\begin{abstract}
  Inference for mechanistic models is challenging because of nonlinear
  interactions between model parameters and a lack of identifiability.
  Here we focus on a specific class of mechanistic models, which we
  term stable differential equations.  The dynamics in these models
  are approximately linear around a stable fixed point of the system.
  We exploit this property to develop fast approximate methods for
  posterior inference.  We illustrate our approach using simulated
  data on a mechanistic neuroscience model with EEG data.  More
  generally, stable differential equation models and the corresponding
  inference methods are useful for analysis of stationary time-series
  data.  Compared to the existing state-of-the art, our methods are
  several orders of magnitude faster, and are particularly suited to
  analysis of long time-series ($>$10,000 time-points) and models of
  moderate dimension (10-50 state variables and 10-50 parameters.)
  \end{abstract}

\section{Introduction}

Differential equations (DEs) are an integral part of modelling in many
different scientific domains. The values of parameters in such models
are key for providing theoretical predictions and interpreting
experimental results. However, often external constraints on parameter
ranges are weak and their values are known only with considerable
uncertainty. From a methodological point of view, two important
challenges for Bayesian inference in Stochastic DE
(SDE) models are: (i) integrating out state variables to obtain the
marginal likelihood for parameter values that drive the system
dynamics; and (ii) designing effective proposals for efficiently
exploring the parameter space.  Pseudo-marginal algorithms, such as
particle Markov chain Monte Carlo (MCMC) are focused primarily on the
first of these challenges \citep{Andrieu2010}, Langevin and
Hamiltonian Monte Carlo algorithms \citep{Girolami2011a, Hoffman2014a}
on the second.

The difficulty of these challenges depends primarily on the following
problem parameters: $n$, the number of time-points in the data-set; $d$,
the dimension of the state-space; and $p$, the number of the unknown
parameters.  We are interested in problems that are difficult due to all of
these, as this is typical for many modern applications of DE Models (DEMs). Attempts to reduce complexity are now often faced
with systems where mathematical simplifications severely reduce the
meaningfulness of the model to the domain scientist.
Existing pseudo-marginal algorithms and Hamiltonian Monte Carlo
methods do not scale well with these problem parameters. The state of
the art methodology for accurate inference in such inference problems
would be to combine these two classes of algorithms
\citep{Dahlin2014}, but this approach still does not scale to the
cases in which we are interested: namely where $d \approx 10$ to $50$,
$p\approx 10$ to $50$, and $n>10^4$.  This regime is common in many
applications, including Computational Neuroscience and Systems Biology
-- see, e.g., \cite{Keener1998, Murray2002, Murray2003,
  Izhikevich2007, ErmenTer2010} and the references therein. For such
applications, one of which may be found in Sec.~\ref{sec:npm},
accurate inference using standard implementations of particle MCMC may
take several months on a desktop computer.

Where ``exact'' inference is not computationally feasible, approximate methods may still be. However, because parameters are often not identifiable in DEMs \citep{Golowasch2002}, methods that solely find point estimates of parameters or Laplace approximations to the posterior are not appropriate (see the discussion of sloppiness in \cite{Girolami2011a}). Approximating the posterior marginals (as in variational, expectation propagation or the integrated nested Laplace methods) is also not appropriate, since the dependency structure in the posterior often is of more interest than the posterior marginals themselves.
In this paper we present an approximate but {\em scalable} alternative. In order for it to be accurate, we need to place restrictions on the allowed dynamical behaviour. Specifically, we consider here only system dynamics in the vicinity of stable equilibrium points. There is a good range of applications where such models are used. Our showcase application is the analysis of an electroencephalogram (EEG) time-series using a mechanistic model. We improve on existing methodology for this application, which can be found in \cite{Moran2009, Pinotsis20121261, Abeysuriya2015}.

The innovations in this paper concern both the
estimation of the marginal likelihood of the parameters, and the
exploration of the parameter space. The spectral density of systems linearized around
a stable equilibrium point can be calculated analytically, rather than by direct numerical solution. Making use of this
fact and the Whittle likelihood, we can reduce the computational
effort required to calculate the marginal likelihood by orders of magnitude. Furthermore,
we introduce heuristics that quantify the accuracy of these approximations.
Exploring the parameter space can be dramatically simplified by a reparameterization that uses equilibrium values as model parameters, which often results in a reduction of computational effort by an additional order of magnitude. Finally, we find derivatives of our Whittle likelihood analytically in order to use gradient-based Monte Carlo methods for efficient exploration.

In the following Sec.~\ref{sec:de_defs}, we describe more precisely
the class of models to which the methods in this paper are directly
applicable.  Section \ref{sec:methodology} introduces the main
methodological innovations.  Section \ref{sec:fhn} then illustrates
the properties of our new approach with the well known FitzHugh-Nagumo
equations \citep{FitzHugh1961,Nagumo1962,Izhikevich2007}. Section
\ref{sec:npm} then applies our method to a more challenging parameter
inference problem from computational neuroscience.

\subsection{Stability and stochasticity in nonlinear dynamical systems}
\label{sec:de_defs}
A general form for many continuous-time nonlinear dynamical systems is
\begin{equation} \label{eq:nonlinear_de}
\frac{d}{d t} \mathbf{x}(t) = \mathbf{F}[\mathbf{x}(t); \theta] +
\mathbf{p}(t; \theta),
\end{equation}

\noindent where $\mathbf{x}$ is a vector of states with length $d$,
$\theta$ is a vector of parameters, and $\mathbf{F}$ is a
vector-valued nonlinear function, also with $d$ components. This form
separates the time-invariant ``system'' given by $\mathbf{F}$ and
$\theta$ from an explicitly time-dependent ``input''
$\mathbf{p}$. Aspects of many natural and engineered systems are
approximated well by this form, due to large differences in the
time-scales of other processes. For example, brain connectivity
changes considerably over a human lifetime, but typically can be
considered as a static structure for the description of neural
processes lasting milliseconds to seconds.

If $\mathbf{p}(t; \theta)$ is a deterministic function, then
Eq.~\eqref{eq:nonlinear_de} represents a system of coupled Ordinary
DEs (ODEs).  This is particularly appropriate if we
model a stimulus (a distinct event, or series of events) through the
input $\mathbf{p}(t; \theta)$.  We then assume that $\mathbf{p}(t;
\theta)$ is zero as default, but becomes non-zero in a characteristic
way at times when the system receives a stimulus.  If $\mathbf{p}(t;
\theta)$ is the time derivative of Brownian motion, then
Eq.~\eqref{eq:nonlinear_de} represents a multivariate diffusion
process with drift term $\mathbf{F}[\mathbf{x}; \theta]dt$.  The
conventional notation for diffusion processes can be obtained by
multiplying through by $dt$, and setting $\mathbf{p}(t; \theta) dt =
\Sigma(\theta)d\mathbf{B}_t$, with $\Sigma(\theta)$ a diagonal matrix,
and $\mathbf{B}_t$ standard uncorrelated multivariate Brownian motion.
We assume that $\mathbf{p}(t; \theta)$ has mean zero. Such an
SDE is particularly appropriate for
``resting state'' conditions, where the system is not primarily driven
by separable and characteristic events (at least to our knowledge),
but noisy internal and/or external fluctuations dominate.

A further distinction that is key for this paper is between what we
refer to as stable and unstable systems. To understand these labels,
we first have to discuss the so-called equilibrium or fixed points of
a system.  A point $\mathbf{x}^*$ is an equilibrium point of the
system dynamics if its state remains constant in this point: $\left.d
\mathbf{x}/d t\right|_{\mathbf{x}=\mathbf{x}^*}=0$. To be more
specific, we consider here equilibrium points for ``least input''
scenarios: in the absence of all stimuli (deterministic case) or for
exactly average input (stochastic case), respectively.  In both cases,
we would set $\mathbf{p}(t;\theta)\equiv 0$ in
Eq.~\eqref{eq:nonlinear_de} due to our definition of the input term
above.  The equilibria are then solutions of $\mathbf{F}[\mathbf{x}^*;
  \theta] = 0$.  The stability of an equilibrium point is evaluated by
examining the eigenvalues of $\mathcal{J}$, the Jacobian matrix of
$\mathbf{F}$ evaluated for $(\mathbf{x}^*;\theta)$.  The condition for
stability of an equilibrium point is
\begin{equation}
\max_{n=1,\ldots,d}( \mathrm{Re}[ \lambda_n ] ) < 0,
\end{equation}

\noindent where $\lambda_1,\ldots,\lambda_d$ are the eigenvalues of
$\mathcal{J}(\mathbf{x}^*;\theta)$.  The Jacobian is calculated by
taking the partial derivatives of $\mathbf{F}$ with respect to the
state variables
\begin{equation}
[\mathcal{J}]_{ij}(\mathbf{x}^*,\theta) = \left.\frac{ \partial
  [\mathbf{F}]_i(\mathbf{x};\theta)}{\partial [\mathbf{x}]_j
}\right|_{\mathbf{x}=\mathbf{x}^*}.
\end{equation}

In the following we will be concerned with methods for
systems evaluated near such a stable equilibrium.  We refer to a
system as ``stable'' in the following if it possesses at least one stable equilibrium
point. There may be some regions of the parameter space where the
system has no stable equilibrium points. However, provided that one
can find a region in the available parameter space where it does, then
we still call this a ``stable'' case -- with the implicit
understanding that our parameter inference is valid
only in a region of parameter space with at least one stable equilibrium.
While our methods are applicable to all such stable systems, the accuracy
of posterior inference depends both on the model and on the magnitude
of the input to the system.  We consider this in more detail in
Sec.~\ref{sec:particle_v_marginal}. Furthermore, we present computationally efficient ways of determining whether the approximation is accurate
for a given problem.

The reparameterization method presented in this paper is applicable to
both stable ODEs and stable SDEs.  A necessary, but not sufficient,
condition for the method to be effective is that the system returns to
its stable equilibrium following transient input.  This is guaranteed
if there is only one stable equilibrium and the system returns to that
equilibrium when initialized from any other point in the state space.
When there is more than one attractor state, then the system will
return to the same equilibrium point if the input / stimulus is
sufficiently small. Spectral analysis using the Whittle likelihood is
applicable to stable SDEs.  The methodology we present for rapidly
calculating the spectral density requires the model to be linearized.
It is therefore only applicable when the input is sufficiently small
for the the linearized model to be a good approximation to the
nonlinear model.  A classic example is the description of a pendulum
as harmonic oscillator, which is valid as long the angle of its swings
are small enough for $\sin\phi\approx\phi$.

The dynamics of
mesoscopic brain oscillations in the resting state, as measured for
example with the electroencephalogram (EEG), are a prominent
case in point
\citep{vanRotterdam1982,Liley2002,Breakspear2002735}. Generally
speaking, many nonlinear systems have operating regimes of interest
where the system dynamics are thought to be (or are designed to be)
quasi-linear, as guaranteed by the local stable manifold theorem
\citep{kelley1967stable}.
Note that linearizing around an equilibrium point is different from
repeatedly linearizing the model around a sequence of distinct points
in the state-space.  In general linearizing around an equilibrium
point will be more approximate because the linearization is only
accurate when the system is close to its equilibrium state, cf.\ Section IV of \cite{Bojak2005}. The
effect of this approximation on posterior inference is examined
empirically in Sec.~\ref{sec:particle_v_marginal}.

\section{Methodology} \label{sec:methodology}

\subsection{Model reparameterization: equilibrium values as parameters}
\label{sec:reparam}

A dynamical system is specified by supplying a set of parameters
$\mathbf{\theta}$. Hence, generally its equilibria $\mathbf{x}^*$ will
be a function of these $\mathbf{\theta}$.  For the form in
Eq.~\eqref{eq:nonlinear_de} discussed above, this follows from solving
$\mathbf{F}[\mathbf{x}^*; \theta] = 0$.  For stable ODEs,
$\mathbf{x}^*$ does not explicitly enter the likelihood calculations.
Nevertheless, the likelihood is implicitly a function of
$\mathbf{x}^*$, and it can be more sensitive to values of
$\mathbf{x}^*$ than to values of $\mathbf{\theta}$.  For stable SDEs,
the likelihood calculations we use depend on linearizing the system
around $\mathbf{x}^*$.  In this case the likelihood is explicitly a
function of $\mathbf{x}^*$.

In this paper we use an MCMC algorithm to explore the posterior on
$\theta$, and find that the posterior may be explored more efficiently
by reparameterizing the model such that $\mathbf{x}^*$ is treated
directly as a parameter of the model. Thus rather than treating
$\mathbf{x}^*$ as a function of $\theta$ (as in the original
parameterization), some elements of $\mathbf{\theta}$ can be calculated
as a function of $\mathbf{x}^*$.  If $\mathbf{\theta}$ is of dimension
$d_{\theta}$ and $\mathbf{x}^*$ of dimension $d_x$, then the combined
parameter space $(\mathbf{\theta},\mathbf{x}^*)$ has dimension
$d_{\theta} + d_x$ but only $d_{\theta}$ degrees of freedom.  One
could choose any combination of $d_{\theta}$ parameters to parameterize
the model.  Then the remaining $d_x$ parameters would be a function of
the chosen $d_{\theta}$ parameters.  We therefore introduce the
subscripts $s$ and $d$ to distinguish between these parameter subsets:
the MCMC is applied to the reparameterized space $\theta_s$, with
$\theta_d$ being a deterministic function of $\theta_s$.

As an example, suppose that the equations for the equilibrium points
were of the simple form
\begin{equation} \label{eq:ss}
\mathbf{F}[\mathbf{x}^*, \mathbf{\theta}_s] + \mathbf{\theta}_d = 0,
\end{equation}

\noindent where $\mathbf{F}$ is a nonlinear function with $d_x$
components, $\mathbf{\theta}_d$ is a vector of parameters also with
$d_x$ elements, and $\mathbf{\theta} = (\mathbf{\theta}_s,
\mathbf{\theta}_d)$.  Then we can simply reparameterize the model in
terms of $(\mathbf{\theta}_s, \mathbf{x}^*)$, and calculate
$\mathbf{\theta}_{d} = - \mathbf{F}[\mathbf{x}^*_s,
  \mathbf{\theta}_{s}]$. Let us compare the form in Eq.~(\ref{eq:ss})
with the one that arises for our systems of interest from
Eq.~\eqref{eq:nonlinear_de}. Note that above we had defined
$\mathbf{p}(t;\theta)\equiv 0$ as default for the input. What if
instead either stimuli or resting state fluctuations occurred on top
of some constant background input
$\overline{\mathbf{p}}+\mathbf{p}(t;\theta)$?  We can bring this into
our standard form by absorbing the constant background input
$\tilde{\mathbf{F}}[\mathbf{x}, \theta]\equiv\mathbf{F}[\mathbf{x},
  \theta]+\overline{\mathbf{p}}$, and using $\tilde{\mathbf{F}}$
instead in Eq.~\eqref{eq:nonlinear_de}.  Then solving for equilibria
of $\tilde{\mathbf{F}}[\mathbf{x}; \theta]=0$ naturally has the form
of Eq.~(\ref{eq:ss}), with the background input
$\overline{\mathbf{p}}$ serving as the $\mathbf{\theta}_d$.  Our
method can also be applied in cases when the dimensions of
$\mathbf{F}$ and $\mathbf{\theta}_d$ do not coincide, i.e., if the
background input is added only to some equations.  It then still may
be possible to obtain closed form expressions for $\theta_d$ in terms
of $\mathbf{x}^*$ and $\theta_s$, as for the example given in
Sec.~\ref{sec:npm}.  Mathematical details of this more involved case
can be found in the supplementary Sec.~\ref{sec:reparam_supp}.

How can such reparameterizations be advantageous?  Likelihood
distributions for DEMs are typically complicated, with strong
dependencies -- linear and/or nonlinear -- between model
parameters. If equilibrium values have weaker nonlinear dependencies
with model parameters than model parameters do with each other then
the posterior will be simplified when the model is reparameterized,.
Consequently, fewer MCMC iterations will be required.  Furthermore,
for a reparameterization in closed form the computational cost per
MCMC iteration is reduced further, since it is then not necessary to
solve a nonlinear system of equations for the equilibrium.

Some issues that arise when designing a sampler on the reparameterized
space are (i) correctly evaluating prior densities and proposal
densities, (ii) ensuring
that the proposal does not frequently propose parameter sets outside
the support of the prior, and (iii) in the case of multiple stable
equilibria, choosing which one to use.  These issues can
all be dealt with in a principled way, cf.\ Sec.~\ref{sec:reparam_supp},
and do not limit the scope of applicability of our methods in practice.

\subsection{The spectral density of stable SDEs}
\label{sec:spec_dens_summ}
If $x^T(t)$ is a component of the multivariate stochastic process
$\mathbf{x}(t)$ measured over time interval $T$, then the (power)
spectral density of $x(t)$ can be estimated with
\begin{equation}
S_{xx}(\nu) = \frac{1}{T}\left|X^T(\nu)\right|^2,
\end{equation}

\noindent where $X^T(\nu)$ denotes the Fourier transform of $x^T(t)$
with ordinary frequency $\nu$. In applications one typically measures
$n$ samples of $x(t)$ during $T$ at regular intervals $\Delta
t=T/n$. The (power) spectral density is then estimated as
\begin{equation}
\frac{1}{\Delta t}S_{xx}\left(\nu=\nu_k\equiv
k/T\right)\quad\longrightarrow\quad S_k=|X_k|^2,
\end{equation}
with $k=0,...,n-1$ and the Discrete Fourier Transform (DFT) $X_k$ of
these samples.

Stable SDEs can be approximated by linearizing around a fixed
point. For the linearized system, one can relate the Fourier transform
of the noise (input) $\mathbf{P}$ to that of the state (output)
$\mathbf{X}$
\begin{equation}
\mathbf{X}(\nu) = \mathcal{T}(\nu)\cdot \mathbf{P}(\nu),
\end{equation}

\noindent where the matrix $\mathcal{T}$ that maps $\mathbf{P}$ to
$\mathbf{X}$ is called the transfer function.  An analytic expression
for $\mathcal{T}$ can be obtained using the eigen-decomposition of
$\mathcal{A}$ \citep{Bojak2005}. We can use these analytic results to
rapidly compute (spectral) model predictions
\begin{equation}
\mathbf{f}(\nu_k)\equiv\frac{1}{\Delta t}\left|\mathcal{T}(\nu_k)\cdot
\mathbf{P}(\nu_k)\right|^2,
\end{equation}

\noindent which can be directly compared with the measured
$S_k$. Analytic derivatives of the modeled spectral density can also
be calculated.  Section \ref{sec:spec_dens} in the supplement contains
a detailed discussion of these estimation methods.  The computational
complexity of the spectral density calculation is $\mathcal{O}(d^3)$.
The computational savings that can result from working in the
frequency-domain as opposed to the time-domain are discussed in
Sec.~\ref{sec:whittle_def}.


\begin{table}[b!]
\caption{MCMC algorithms used to generate results in Secs.~\ref{sec:fhn} and \ref{sec:npm}} \label{tab:algs}
\begin{center}
\begin{tabular}{ |l|l|l| }
  \hline Algorithm Name & Applicability & Likelihood \\
  \hline MwG &  nonlinear ODEs & Eq.~\eqref{eq:dl} \\
  Particle Marginal MwG & nonlinear SDEs & MC of Eq.~\eqref{eq:ml} \\
 Kalman Marginal MwG & linearized SDEs & Eq.~\eqref{eq:ml}\\
 Whittle smMALA & linearized SDEs & Eq.~\eqref{eq:whittle} \\ \hline
\end{tabular}
\end{center}
\end{table}

\subsection{Whittle likelihood for SDEs}\label{sec:Whittle}

In this section we describe our novel methodology for
inference in SDEs with a stable equilibrium point. We use an MCMC
framework, where calculating the likelihood of the parameters requires
integrating over the state space.  The resulting marginal likelihood
estimate enters into the acceptance ratio of MCMC algorithms for
parameter estimation.  Details of the MCMC algorithms
(Metropolis-within-Gibbs, MwG, and simplified manifold MALA, smMALA)
used in this paper can be found in the supplementary
Sec.~\ref{sec:mcmc_alg}.  In Tab.~\ref{tab:algs} we list the variants
of MwG and smMALA used to generate our results.

The novelty in our approach lies in
the way that the contributions in Secs.~\ref{sec:reparam} and
\ref{sec:spec_dens_summ} are used within existing MCMC
methodology.  We will apply the Whittle likelihood
\citep{whittle1962gaussian} to stable DEMs of the form
\begin{align}
\mathrm{d}\mathbf{x}(t) &= \mathrm{F}\left[\mathbf{x}(t);\theta\right]\mathrm{d}t +
\Sigma(\theta) \mathrm{d}\mathbf{B}_t \label{eq:model1} \\
\mathbf{y}_i &= \mathrm{L}(\mathbf{x}_i; \theta) +
\mathbf{n}_i(\theta) \label{eq:model2}
\end{align}
\noindent where $\mathbf{y}_i$ is a vector of observations at the
discrete time index $i$ of the system state $\mathbf{x}_i\equiv
\mathbf{x}(t=t_i)$, $\mathrm{L}$ is a linear function, and
$\mathbf{n}_i$ is a discrete-time process for which it is possible to
compute the spectral density, the simplest example of which is white
noise.  The parameter set, $\theta$, now includes parameters that
influence the observation process, $\mathbf{n}_i(\theta)$, as well as
$\mathrm{F}[\mathbf{x}(t);\theta]$ and $\Sigma(\theta)$.

The Whittle likelihood approximates the marginal likelihood,
\begin{equation} \label{eq:ml}
p_{\theta}(y_{1:n}) = \int p_{\theta}(y_{1:n} | x_{1:n})  p_{\theta}(x_{1:n})\, dx_{1:n}.
\end{equation}
In Sec.~\ref{sec:whittle_def} of the Supplement we summarize how this
likelihood can expressed purely in terms of the theoretical frequency
representation $\tilde{f}(\nu_k;\theta)$, depending on model
parameters $\theta$, and the empirical one $\tilde{S}_k$ derived from
data, without referring at all to the time domain:
\begin{equation} \label{eq:whittle}
p_{\theta}(y_0,\ldots,y_{n-1}) =
p_{\theta}\left(\tilde{S}_0,\ldots,\tilde{S}_{n-1}\right) \approx
\prod_{k=1}^{n/2-1} \frac{1}{\tilde{f}(\nu_k;\theta)}
\exp\left[-\frac{\tilde{S}_k}{\tilde{f}(\nu_k;\theta)}\right].
\end{equation}
Both frequency representations take into account
Eq.~(\ref{eq:model2}), e.g., if only white noise is added then
$\tilde{f}(\nu_k;\theta)=f(\nu_k;\theta)+\mathrm{const.}$

Linearization of the model around a stable equilibrium point
(described in Sec.~\ref{sec:spec_dens}) yields a stationary
approximate model for which the spectral density exists, and which we
may differentiate analytically. Our main reason for using the Whittle
likelihood is that we may exploit a computational saving in the case
of evaluating the spectral density of a partially observed SDE,
cf.\ supplementary Sec.~\ref{sec:spec_dens}.  In the cases
considered in this paper in which only a single variable is observed,
evaluating the spectral density at $n$ points (as we do in the Whittle
likelihood) has complexity $\mathcal{O}(n d)$, which compares
favourably to the $\mathcal{O}(n d^3)$ complexity of the Kalman
filter. We further note that the use of the Whittle likelihood may be
intuitive to practitioners, who are often familiar with a frequency
domain representations of their data. For example, quantitative
studies of the electroencephalogram (EEG) often make use of the
(power) spectral density.

The approximation made by the Whittle likelihood is two-fold.  First,
the model is linearized at a stable fixed point of the system.
Second, the likelihood of the linearized model is approximated in the
frequency domain by Eq.~\eqref{eq:whittle}.  See Sec.~\ref{sec:whittle_def} in the
Supplementary for more details.  Both of these
approximation errors can be quantified, as we show in
Sec.~\ref{sec:particle_v_marginal} and supplementary
Sec.~\ref{sec:kalman_v_whittle}.  The linear approximation is well
suited to SDEs operating close to a stable equilibrium as the dynamics
are approximately linear in this case.  When this condition is not
satisfied, a more appropriate choice of likelihood estimate is the
particle filter \citep{Gordon1993}, an importance-sampling Monte Carlo
method that yields an unbiased approximation of the marginal
likelihood up to discretization error.  The particle filter approach
is often accurate, but is usually computationally expensive.  See
Sec.~\ref{sec:pf_supp} in the Supplementary for more details.

For linear models the marginal likelihood can be evaluated exactly
using the Kalman filter.  By numerical
experimentation, cf.\ Sec.~\ref{sec:kalman_v_whittle}
in the Supplement, we have found that the
Kalman filter and Whittle likelihood posteriors are similar when
\begin{align}
\label{eq:heuristic}
\frac{\phi}{n} &< 0.01 \max_k\left[ f(\nu_k)\right],\\
\label{eq:phi}
\phi&\equiv\sum_{h=-\infty}^{\infty} |h|\left|\gamma_{xx}(h\cdot\Delta
t)\right|,
\end{align}
where $\gamma_{xx}(\tau)$ is the autocovariance function.
This is illustrated in Fig.~\ref{fig:2} in the Supplement. Since it is
quicker to compute this than to do a full
comparison between posterior distributions, we recommend using
Eq.~\eqref{eq:heuristic} as a heuristic rule to decide whether the
Whittle likelihood is a sufficiently accurate approximation to the
true likelihood.  For the example in Sec.~\ref{sec:npm}, estimating the
posterior distribution using the Kalman filter is intractable and we
rely on the heuristic in Eq.~\eqref{eq:heuristic} to give an
indication of whether the Whittle likelihood is accurate.

\section{Analysis using the FitzHugh-Nagumo equations}
\label{sec:fhn}

The FitzHugh-Nagumo (FHN) model
\citep{FitzHugh1961,Nagumo1962,Izhikevich2007} is a highly simplified
description of neural activity and exhibits nonlinear oscillations.
These equations have been used before to test methodology for general
nonlinear problems \citep{Girolami2011a, Jensen2012}.  In this section,
we use the FHN system to illustrate the properties of our approach,
comparing it with existing ones. We begin in Sec.~\ref{sec:FHNdummy}
(details in the supplementary Sec.~\ref{sec:det_fhn}) by
using a deterministic FHN variant in order to
illustrate the improvement in efficiency of Metropolis-within-Gibbs
when run on a reparameterized model. Next in Sec.~\ref{sec:particle_v_marginal},
we study a stochastic version of the
FHN model in order to compare the use of the three time domain
likelihoods described in Secs.~\ref{sec:whittle_def} and \ref{sec:like_supp}, namely the
particle filter, the extended Kalman filter, and the Kalman filter
with a time-homogeneous linearization of FHN about a stable
equilibrium point.

There are parameter sets for which the FHN model has a stable
equilibrium point, and where, given a sufficiently large perturbation
away from the stable equilibrium, the model produces a nonlinear
transient oscillation.  There are also parameter sets where a
sufficiently large perturbation from stable equilibrium can cause the
the system to move towards a limit-cycle attractor.  However, we do
not consider that case here. The FHN equations that we will use here
are
\begin{align}
\frac{d}{dt} V(t) &= V(t) [a - V(t)] [V(t) - 1] - w(t) + I_0 +
I(t) \label{eq:fhn1}\\ \frac{d}{dt} w(t)&= b V(t) - c w(t) + d +
P(t).\label{eq:fhn2}
\end{align}
We assume the following observation model,
\begin{equation}
y_i = V(t_i=i\cdot\Delta t) + n_i
\end{equation}
\noindent where $\Delta t$ is some constant time-step and the $n_i$ are
i.i.d.\ normal random variables, for $i=0,\ldots,n-1$.

The state variables $V$ and $w$ are referred to as the membrane
potential and the recovery variable, respectively, when the model is
used in the neuroscience domain.  Compared to the standard FHN form
\citep{FitzHugh1961,Nagumo1962,Izhikevich2007}, we have added here
terms $d$ and $P(t)$ to the recovery variable, to allow additional
analysis below. However, this is not intended as a new model proposal
in the domain science, and obviously the standard form will be
recovered if these terms are set to zero.  Furthermore, as discussed
in Secs.~\ref{sec:de_defs} and \ref{sec:reparam}, we intend $P(t)$
to be zero by default in the ODE and mean-zero in the SDE case, with
$d$ parameterizing any constant perturbation. Similarly, we consider
$I(t)$ here as zero by default / mean-zero, and hence in comparison to
the standard notation have explicitly separated out a possible
constant input as $I_0$.

Here and in the following we always consider equilibria for default
inputs $P(t)=I(t)=0$.  The steady-state equations, also known as
nullclines, are then obtained by setting the derivatives $dV/dt$ and
$dw/dt$ to zero.  If constant inputs $d$ and $I_0$ are absent as well,
then it is easy to see from Eqs.~\eqref{eq:fhn1} and \eqref{eq:fhn2}
that there is an equilibrium at $V^*=w^*=0$. As shown, equilibrium
values of state variables will be indicated by a star
superscript. This particular equilibrium will be stable if $a+c>0$ and
and $ac+b>0$, as follows from computing the eigenvalues of the
Jacobian:
\begin{equation}\label{eq:fhnj}
\mathcal{J} =
 \begin{pmatrix}
  -a & -1 \\ b & -c \\
 \end{pmatrix}.
\end{equation}

\noindent There are multiple ways to reparameterize the model, and the
most appropriate choice depends on the inference problem.  If both $V$
and $w$ were observed, and the aim was to infer $(a, b, c, d, I_0)$,
we would update $V^*$ and $w^*$ using the MCMC proposal and then,
given the values of $V^*$ and $w^*$, solve for the inputs $d$ and
$I_0$:
\begin{align}
I_0 &= - V^* ( a - V^*) ( V^* - 1) + w^* \label{eq:fhnss1} \\ d &= -b
V^* + c w^* \label{eq:fhnss2}
\end{align}
In the MCMC algorithm for the reparameterized model, the updates for
$(a, b, c, V^*, w^*)$ would be stochastic with a Metropolis-Hastings
kernel.  The updates for $(I_0, d)$ would be deterministic using
Eqs.~\eqref{eq:fhnss1} and \eqref{eq:fhnss2}.  Alternatively, if only
$V$ was observed and $d$ was known we would update only $V^*$ using the
MCMC proposal since the likelihood is no longer sensitive to the value
of $w^*$.  In this case the value of $w^*$ is determined through
the other parameters, and we solve for $I_0$:
\begin{align}
w^* &= \frac{1}{c} ( b V^* + d ) \label{eq:fhnss3} \\ I_0 &= - V^* ( a
- V^*) ( V^* - 1) + w^* \label{eq:fhnss4}
\end{align}

When the model is not reparameterized, the equilibria can be obtained by
solving a cubic equation for $V$, and then substituting these values
into a linear equation for $w$.  The coefficients of the cubic
$V\/$-nullcline depend on all of the parameters, hence the
steady-states are then nonlinear functions of the model parameters.
For a given set of parameters, the system can have more than one
stable equilibrium.  We define a unique mapping from parameters to
equilibrium by selecting the steady-state $V$-component that is
closest to the mean of the $V\/$-observations.

There are some scenarios where the model parameters of the FHN model
are not identifiable from the data.  In results for the FHN model, we
consider a scenario where the model parameters are identifiable as
this eases visualisation and interpretation.  In the more challenging
scenario where the parameters are not identifiable from the data we
recommend using the prior distribution to stabilize the variance of
the posterior distribution.  Our analysis of a Neural Population Model
in Sec.~\ref{sec:npm} is an example of this.

\subsection{Results for deterministic FitzHugh-Nagumo equations}
\label{sec:FHNdummy}
From Sec.~\ref{sec:particle_v_marginal} onwards, we focus on stable
SDEs, where both the methodology on model reparameterization and
likelihood approximation can be applied.  Here we briefly summarize
the results of applying the reparameterization methodology to the
deterministic FitzHugh-Nagumo equations.  For certain parameter sets
this model has a stable fixed point, and is therefore a stable ODE.
As measured by differences in the ESS, the Metropolis-within-Gibbs
algorithm is faster by a factor of around 14 when the model is
reparameterized, as shown in Tab.~\ref{tab:ess}.  Further details on
the model parameters and MCMC samples are given in
Sec.~\ref{sec:det_fhn} of the Supplement.
\begin{table}
\begin{center}
%
\begin{tabular}{|c|c|c|c|c|}
 \hline & \multicolumn{2}{|c|}{Original parameterization} &
 \multicolumn{2}{|c|}{Steady-state reparameterization} \\ \hline
 Parameter & Proposal s.d. & ESS & Proposal s.d. & ESS \\ \hline $a$ &
 5 & 13 & 3.5 & 175 \\ $b$ & 250 & 22 & 250 & 144 \\ $c$ & 1 & 10 & 3
 & 174 \\ $I_0$ & 2.5 & 9 & n/a & 249 \\ \hline & Average ESS &
 \textbf{13.5} & Average ESS & \textbf{185.5} \\ \hline
\end{tabular}
\end{center}
\caption{Results for Metropolis-within-Gibbs MCMC algorithm with and
  without reparameterization.  Total number of MCMC iterations was
  100,000.  ESS was calculated on the second half of the samples
  (i.e., the last 50,000 samples) to remove the effects of burn-in.
  Proposal standard deviation (s.d.) for $V^*$ in the reparameterized
  algorithm was 0.015.} \label{tab:ess}
\end{table}

\subsection{Evaluating posterior accuracy under linearized model}
\label{sec:particle_v_marginal}

As discussed in Sec.~\ref{sec:whittle_def}, the Whittle likelihood
approximates the marginal likelihood, $p_{\theta}(y_{1:n})$.  The
purpose of this section is to examine the accuracy of this
approximation in more detail. We first look at errors arising from
linearizing the model, using the stochastic FitzHugh-Nagumo equations
as an example.  We are interested in cases when the dynamics
of the nonlinear model are approximately linear around some stable
equilibrium.  As discussed in supplementary Sec.~\ref{sec:spec_dens},
the accuracy in the approximation depends on the
magnitude of the input to the system, as this determines the amplitude
of the system dynamics.  For a given set of system parameters, the
linearization can be made arbitrarily accurate by choosing a small
enough magnitude for the input.  However, this is not a strong enough
result to guarantee accurate posterior inference.  Suppose that the
linearization is accurate for the true system parameters and the true
input parameters.  There may be regions of the parameter space where
the likelihood is high under the linearized model but the
linearization is inaccurate.  For example, if a dynamical system has
two stable fixed points and the magnitude of the noise that feeds into
the system is sufficiently large, the system can transition between
approximately linear dynamics around one fixed point to approximately
linear dynamics around the other fixed point.  In the linearized
model, the system always stays relatively close to the stable
equilibrium point that the system was linearized around.  It is not
uncommon in dynamical systems models for the behaviour of the system
to be different in different regions of the parameter space.  Hence,
for one parameter set the dynamics of the nonlinear model are
approximately linear around a single fixed point, but for a different
parameter set the system switches between two different fixed points.

\begin{figure}[htbp]
\floatpagestyle{empty}
\centering \includegraphics[width=\textwidth]{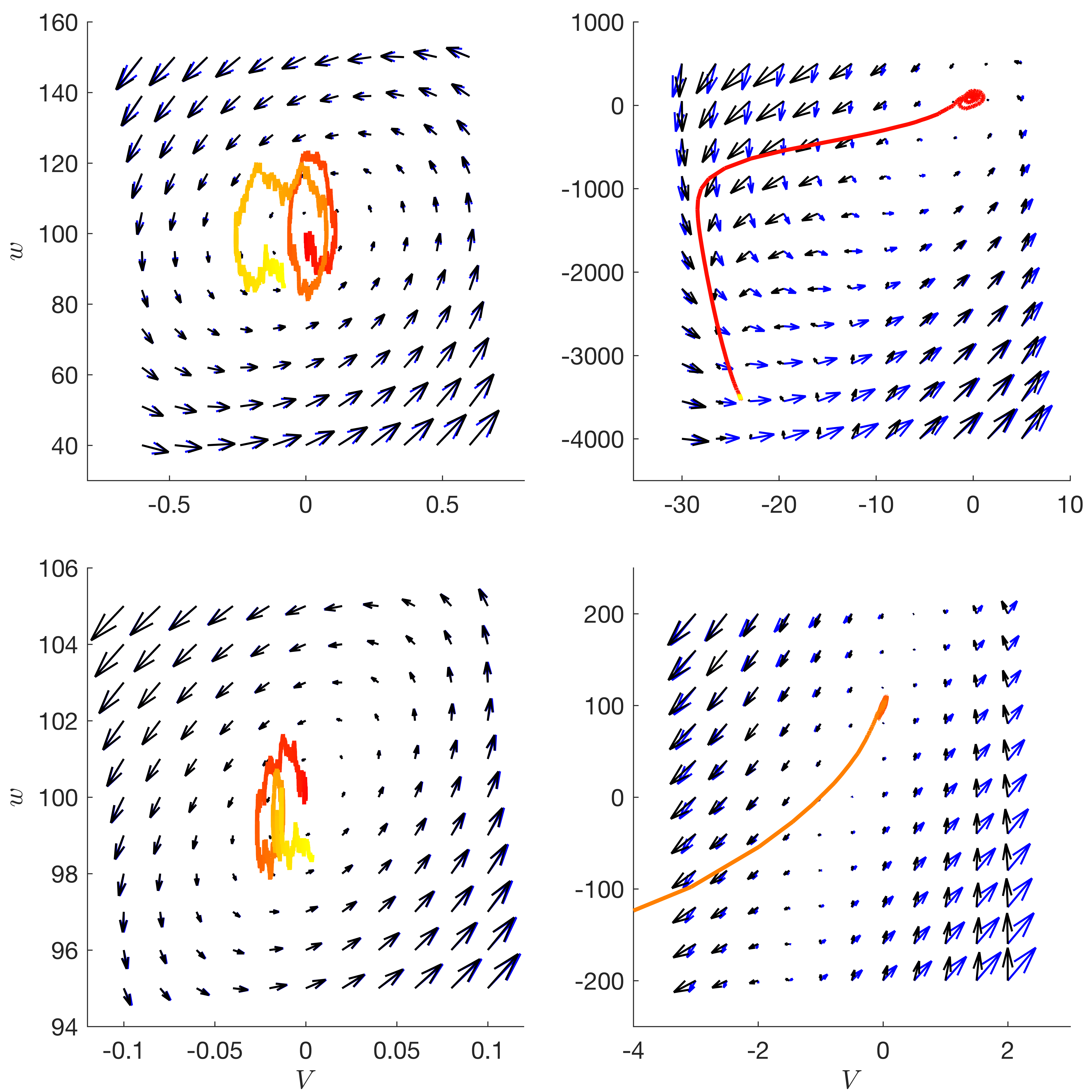}
\caption{ Vector fields for the FitzHugh-Nagumo model (black arrows)
  and its linearization around a stable equilibrium (blue arrows), as
  well as sample path of nonlinear model (coloured line).  Arrows
  represent the drift term, i.e., the deterministic time-derivative.
  \textit{Left column:} Nonlinear model is well approximated by
  linearization.  \textit{Right column:} Linearized model is not a
  good approximation in spite of similar likelihood compared to the
  cases on the left, cf.\ see Figs.~\ref{fig:sfhn2} and
  \ref{fig:sfhn3}. $I(t)=0$ and $P(t)$ was white noise with variance
  $\sigma_{in}^2$. \textit{Top left} parameter values for
  $(a,b,c,d,I,\sigma_{in})$: $(-5, 6000, 40, 4000, 100,
  100)$. \textit{Top right}: $(-30, 6000, 40, 4000, 100,
  100)$. \textit{Bottom left}: $(-30, 6000, 40, 4000, 100,
  10)$. \textit{Bottom right}: $(-150, 3000, 170, 17000, 100, 10)$. }
\label{fig:sfhn1}       
\end{figure}

\clearpage

This is illustrated using a stochastic FitzHugh-Nagumo model in
Fig.~\ref{fig:sfhn1}. It shows in the {\em top left} panel a case with
only one equilibrium point where the linearization works well. In the
{\em top right} panel we see that under the change of just one
parameter linearization and nonlinear model cease to agree well: the
nonlinear model escapes to a different equilibrium point, which is not
predicted by the linearization. In the {\em bottom left} panel we see
that reducing the input noise by a factor of ten stabilizes the
agreement again. However, the {\em bottom right} panel demonstrates
that one can change more parameters to remove agreement again even at
these low input noise levels. It should be noted however that if we
had linearized around the second equilibrium point in
Fig.~\ref{fig:sfhn1} {\em top right} -- the one to which the nonlinear
model escapes -- then we would have found agreement even at the high
noise levels. Clearly the appropriateness of the linearization in a
situation with multiple equilibria has to be confirmed on a case by
case basis.

In order to quantify the error caused by the linearization we ran both
MCMC using a Kalman filter to estimate the likelihood (i.e. marginal
MCMC), and particle MCMC, on data simulated from the FitzHugh-Nagumo
model.  We found the posterior distribution was noticeably broader in
the marginal MCMC samples, i.e., in the algorithm where the model was
linearized (results not shown). This discrepancy necessitated further
empirical investigation, however this was constrained by the fact that
the computational cost of running particle MCMC on the FitzHugh Nagumo
model was very high (requiring more than one month of computing time
on a cluster of 20 cores). This can be reduced, for example by using a
low-level language, e.g., C, a more sophisticated variant of particle
MCMC, e.g., a proposal that makes use of the next data-point, or by
implementation on a computing environment where greater parallelism is
possible, e.g., GPUs.  All of these options are (currently) expensive
in terms of user-time, thus were not pursued here.

As an alternative
to comparing with particle MCMC, the following checks are useful for
partially quantifying the error that comes from linearizing the model:
(i).\ Compare marginal likelihood estimates from the basic Kalman
  filter (i.e. linearizing about the stable point), Extended Kalman
  Filter (EKF), and particle filter on a 1-D grid for each individual
  parameter.
(ii).\ Compare the posterior distribution obtained from running
  marginal MCMC with the Kalman filter to marginal MCMC with the EKF.
(iii).\ Simulate from the linearized model and the nonlinear model at
  parameter sets sampled by marginal MCMC to check for nonlinear
  dynamics.
These require much less computing time than running particle MCMC, and
are also relatively easy to implement. We performed these checks for
several scenarios with the aim of investigating the effect of our
approximations.

\begin{figure}[htbp]
\centering
  \includegraphics[width=\textwidth]{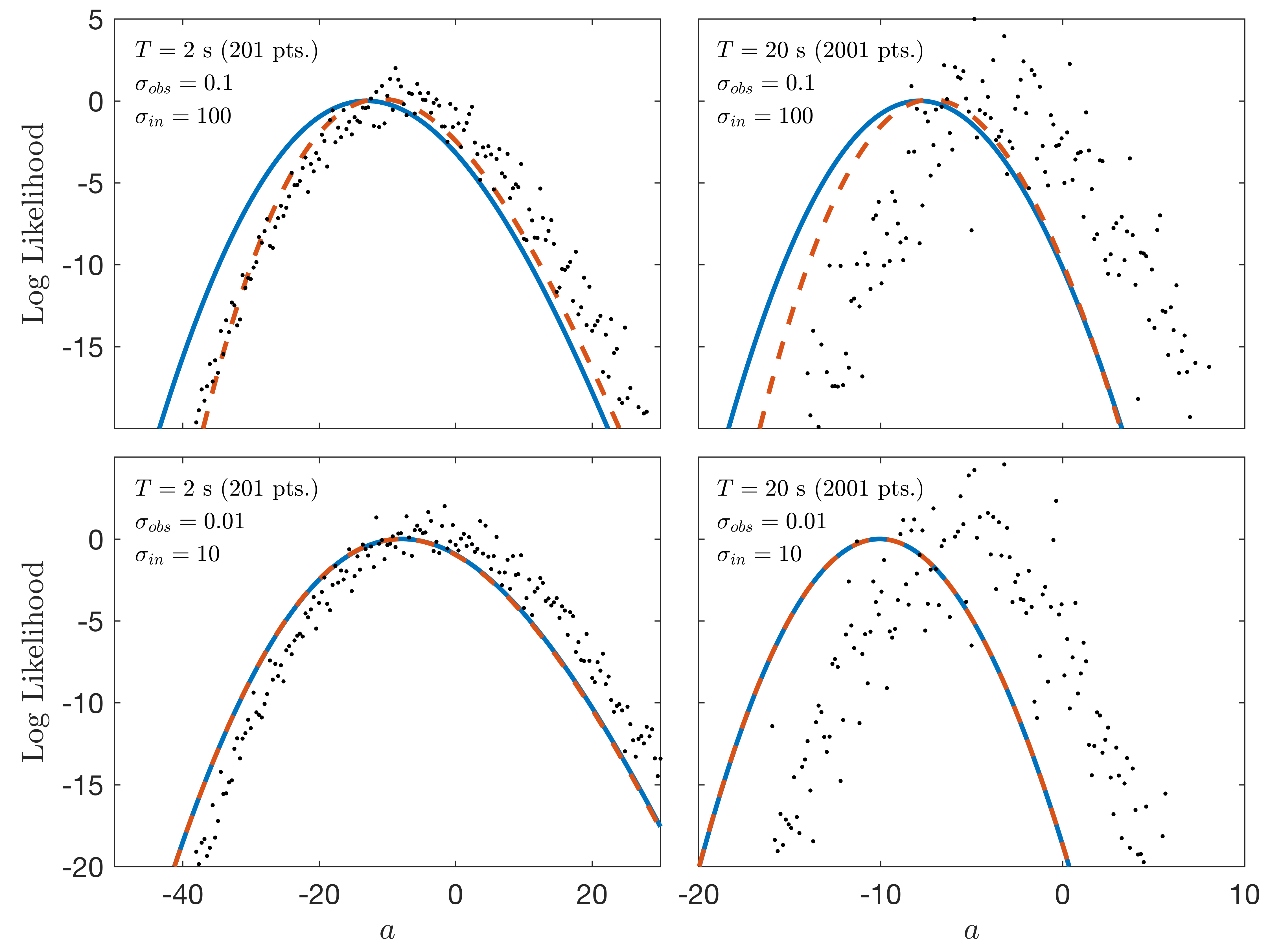}
\caption{Comparison between Kalman filter (solid blue line), EKF (dashed red line) and particle filter (black dots) for the stochastic FitzHugh-Nagumo model. $P(t)$ is white noise with variance $\sigma_{in}^2$, observation noise with variance $\sigma_{out}^2$ is added.  The only unknown parameter is $a$. The marginal likelihood is estimated on a uniformly spaced sequence of 200 $a$ values. Fixed parameters:  $b=6000$, $c=40$, $d=4000$, $I_0=100$, $I(t) \equiv 0$. Time-step in solver and Kalman filter = $10^{-3}$, in observations = $10^{-2}$. Number of particles = $1000$.}
\label{fig:sfhn2}       
\end{figure}

\begin{figure}[htbp]
\centering
  \includegraphics[width=\textwidth]{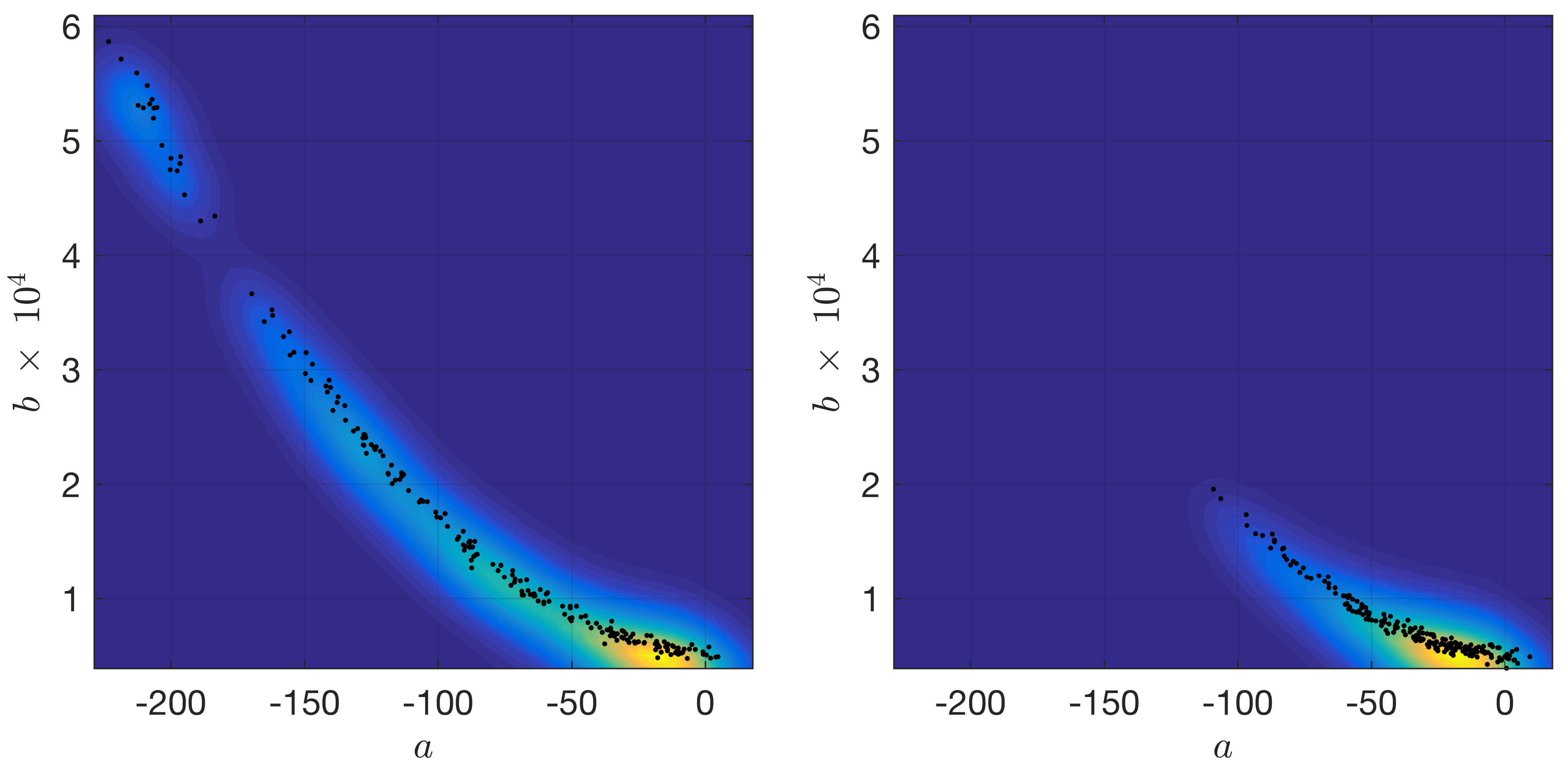}
\caption{Comparison between marginal MCMC with Kalman filter (left) and EKF (right) on a problem with unknown parameters ($a$, $b$, $c$, $I_0$).  Plots show MCMC samples from the joint $(a,b)$ marginal distribution only. Parameter values: $d=4000$, $T=2$, $\sigma_{in} = 10$, $\sigma_{obs} = 0.01$.  Time-step in solver and Kalman filter = $10^{-3}$, in observations = $10^{-2}$.  The MCMC chain started at the true parameters and ran for 10,000 iterations. Plots shows every 500th sample (black dots) and a smoothed histogram (coloured background). }
\label{fig:sfhn3}       
\end{figure}

In summary, when performing linearization around a fixed point we have
observed the following useful properties:
(i).\ There is only a small difference using the Kalman filter (in
  which the linearization is about a stable point) and the EKF when
  the input noise is small - Figure \ref{fig:sfhn2}.
(ii).\ As more data is collected, the posterior variance decreases at
  approximately the same rate with the approximate and exact methods -
  Figure \ref{fig:sfhn2}.
(iii).\ Our approximate posterior distribution tends to accurately
  capture the qualitative dependency structure between parameters -
  Figure \ref{fig:sfhn3}.
(iv).\ Where the likelihood differs between the Kalman filter and the
  EKF, we find that our approach tends to over-estimate uncertainty -
  Figure \ref{fig:sfhn3}.  This makes our approximation suitable for
  embedding within an exact approach to speed it up (e.g., using
  importance sampling or delayed acceptance \citep{Golightly2014a}).

Figure \ref{fig:sfhn2} shows that the error in the posterior is lower
when the input noise and the observation noise are smaller.  (We
decrease both the noise parameters so that the signal to noise ratio
does not change.)  More specifically the Kalman filter and the EKF
results are almost identical in the low noise setting, at least for
the one-dimensional problem.  The level of input noise controls the
component of the error that comes from linearizing the model around a
single point in the state-space, as opposed to linearizing locally
around current state estimates (as in the EKF).  The residual error,
which exists when any form of linearization is done, is found by
comparing the EKF with the particle filter.  As is well known from
previous studies, the EKF produces biased parameter
estimates. Measured relative to posterior variance, the bias increases
as more data is collected, see Figure \ref{fig:sfhn2}.  However,
measured relative to the true parameter value, the bias in the
posterior mean is small. Figure \ref{fig:sfhn2} also shows that
larger data-sets result in larger variances for particle filter
estimates for a fixed number of particles, an issue that is discussed
in \cite{Kantas2014a}.  This reinforces the message in Sec.~\ref{sec:Whittle}
that the number of particles needed to effectively
apply particle MCMC makes the algorithm intractable for many problems
of practical interest. Figure \ref{fig:sfhn3} shows that accuracy on
one-dimensional problems does not necessarily imply accuracy in
higher-dimensional problems with the same model.  The true parameters
in this case correspond to the bottom-left panel in Figure
\ref{fig:sfhn2}, i.e., the input noise is sufficiently low that the
Kalman filter and EKF are almost identical in one dimension.  For the
higher dimensional problem (where $a$, $b$, $c$, $I_0$ are all
unknown), we see that the Kalman filter over-estimates uncertainty,
although it still captures the qualitative dependencies in the
posterior.

\section{Analysis using Neural Population Model}
\label{sec:npm}

Here we consider a problem from neuroscience that is extremely
computationally expensive using particle MCMC, and where application
of the Whittle likelihood makes a significant difference to
computational efficiency compared to the Kalman filter. The use of the
Whittle likelihood, enabled by linearizing about the fixed point,
leads to a reduction in computational cost of a factor of 100 compared
to the Kalman filter and several more orders of magnitude compared to
particle MCMC. This approximation yields an algorithm that runs in
several hours, compared to several months when using the EKF instead.

Other methods based on spectral analysis, similar to the Whittle
likelihood, are currently used in EEG data analysis \citep{Moran2009,
  Pinotsis20121261, Abeysuriya2015}.  These methods have the same
computational complexity as the Whittle likelihood.  The reason for
choosing the Whittle likelihood over these alternative spectral
analysis methods is the theoretical support outlined in Sec.~\ref{sec:whittle_def}.
This makes it possible to analyse the expected
accuracy of the Whittle likelihood.  In particular, we are able to
estimate how long the time-series should be in order to make the bias
introduced by the Whittle likelihood negligible.  As far as we are
aware, this type of analysis is absent from the neuroscience
methodology papers that combine modelling and spectral analysis.  We
also test whether the model reparameterization makes a difference when
applied within the smMALA algorithm (Algorithm \ref{alg:mmala} in the
Supplement).  We find a similar improvement in MCMC efficiency as when
the reparameterization was applied within the MwG algorithm for the
FitzHugh-Nagumo equations in Sec.~\ref{sec:FHNdummy}.

In Secs.~\ref{sec:modeldef} and \ref{sec:npm_reparam} of the
Supplement, we describe in detail the Neural Population Model (NPM)
that we would like to infer the parameters of, and explain how the
reparameterization method can be applied to this model. For reference,
the model employed here consists of the following DEs, where $l=e,i$ for excitatory and inhibitory contributions:
\begin{align}
\tau_l \frac{d}{dt} h_l(t) &= h_l^r - h_l(t) + \sum_{k=e,i}\frac{h_{kl}^{eq} - h_{l}(t)}{|h_{kl}^{eq} - h_{l}^r|} I_{kl}(t),\\
\left( \frac{1}{\gamma_{el}}\frac{d}{dt} + 1 \right)^2 I_{el}(t) &=
q_{el} \left\{N^{\beta}_{el} S_e(t) + \Phi_{el}(t) + \bar{p}_{el} + \delta_{el}\  p(t)\right\}, \\
\left( \frac{1}{\gamma_{il}}\frac{d}{dt} +  1\right)^2 I_{il}(t) &=
q_{il}  N^{\beta}_{il} S_i(t),\\
\left(\frac{1}{v\Lambda}\frac{\partial}{\partial t} + 1\right)^2 \Phi_{el}(t) &= N^{\alpha}_{el} S_e(t),\\
S_k(t) &= S_k^{max}/\left\{1 + \exp\left[
\frac{\bar{\mu}_k - h_k(t)}{\hat{\sigma}_k/\sqrt 2} \right]\right\},
\end{align}
where the Kronecker delta $\delta_{el}$ admits only excitatory noise
input $p$ to this SDE system.

In modeling the EEG with this NPM, one typically assumes that
the EEG observations are linearly proportional to $h_e$ with
some added observational noise.
We hence assume the following observation model,
\begin{equation}
y_i = h_e(i\cdot\Delta t) + n_i
\end{equation}
\noindent where $\Delta t$ is some constant time-step and the $n_i$ are
i.i.d.\ normal random variables, for $i=0,\ldots,n-1$. Figure~\ref{fig:spec}
in the Supplement shows typical pseudo-data and spectral densities
obtained with this model.
Here we will be interested in the
parameters governing the local reaction to incoming synaptic inputs
($\gamma_{ee}$, $\gamma_{ei}$, $\gamma_{ie}$, $\gamma_{ii}$, $q_{ee}$,
$q_{ei}$, $q_{ie}$, and $q_{ii}$), which are commonly influenced by
psychoactive substances and thus of considerable practical importance.
We will assume the other parameter values are known, and simulate data
from the model with a specific parameter set in order to test our
statistical methodology, i.e., we will rely on pseudo-data for the
following analysis.

The NPM introduced above has several different dynamical regimes,
e.g., quasi-linear around a stable fixed point, limit cycles, or
chaotic.  Our working assumption is that if the data is weakly
stationary, the underlying process can be modelled with parameter
values leading to approximately linear dynamics around a stable fixed
point for this NPM.  See Sec.~\ref{sec:particle_v_marginal} for a
detailed discussion of the effect of linearization.  For the
FitzHugh-Nagumo model we found that using the marginal likelihood
calculated with the linearized model resulted in a posterior
distribution that over-estimated uncertainty.  The same may be true
for the NPM, although it is more difficult to test because of the
greater computational resources required to run the Kalman filter,
extended Kalman filter and particle filter.  We have simulated the
model at parameter sets sampled by the MCMC algorithm and found a good
agreement between nonlinear and linearized behaviour, consistent with
the results in \cite{Bojak2005}.


\subsection{Results}
\label{sec:npm_res}

For this problem, particle MCMC is not computationally tractable, so
we cannot make a direct comparison of the different likelihood
choices.  Below we present MCMC results for parameter estimation in
the NPM with approximately linear dynamics.  We use the Whittle
likelihood and the smMALA algorithm (Algorithm \ref{alg:mmala}).  Here
the Whittle likelihood is over 100 times faster than the Kalman
filter.  This is so because there is only one non-zero input, and we
only observe one system state, so the likelihood calculation is faster
by a factor of $O(d^2)$ with here $d=14$.  Following the discussion in
Sec.~\ref{sec:kalman_v_whittle}, we use the definition of $\phi$ in
Eq.~\eqref{eq:phi} and the heuristic in Eq.~\eqref{eq:heuristic} to
determine the minimum time series length for which we expect the
Whittle likelihood to be accurate.  For this model, the theoretical
spectral density $f(\nu_k;\theta)$ for parameters $\theta$ is
calculated using the method in Sec.~\ref{sec:spec_dens} and the
autocovariance function can be calculated numerically
\begin{equation} \label{eq:gamma_f}
\gamma_{xx}(h\cdot\Delta t)=\sum_{k=0}^{n-1} f(\nu_k;\theta) e^{i2\pi
  kh/n}.
\end{equation}
Assuming the parameters of Tab.~\ref{tab:npm} in the Supplement, the
result of this analysis is a minimum time series length of around
$7,000$ time-points, which is equivalent to around 14~s of data.  For
this particular model we found that the smMALA algorithm had better
convergence than the MwG algorithm (Algorithm \ref{alg:mwg}).  We ran
the MwG algorithm for a long time from a random initialisation and it
did not sample parameter values close to the true parameter set
(results not shown).

We chose a product of independent log-normal distributions as the
prior distribution for the model parameters. See Sec.~\ref{sec:npm_prior}
in the supplementary material for more details.
The smMALA algorithm was run on data simulated from a known parameter
set, referred to as the true parameters.  The parameter set used to
initialize the MCMC was generated by random sampling with a procedure
that was blind to the true parameter values.  The algorithm parameters
were as follows: Langevin step-size $h=0.3$, number of MCMC iterations
$=100,000$ and burn-in length $=20,000$.  Table \ref{tab:comparison2}
compares the performance of the smMALA algorithm with the original
parameterization against the performance with the reparameterization.
The reparameterized model has a lower computational cost per MCMC
iteration, and a higher ESS.  The cost per iteration is lower because
the parameters that are updated deterministically in the
reparameterization are explicit function of the stochastically updated
parameters.  In the original parameterisation the steady-state were
defined implicitly, and a nonlinear solver was required to obtain
their values given the other parameters.  Reparameterizing the model
may also have reduced nonlinearities in the likelihood function,
leading to the higher ESS.  The overall gain in efficiency is around a
factor ten for Langevin step-size $h=0.3$.
Table \ref{tab:comparison2} also shows the results for a smaller value
of the step-size parameter, $h$, in smMALA.  In this case the two
parameterizations have similar acceptance rates.  An explanation for
the difference in results for different $h$ values is that for the
higher $h$ value the discretization of the Langevin diffusion is
unstable in a large region of the parameter space for the original
parameterization, but not for the reparameterized model.  In contrast,
for the lower $h$ value the discretization is stable over most of the
parameter space in both parameterisations of the model.

\begin{figure}[htbp]
\floatpagestyle{empty}
  \centering \includegraphics[width=0.8\textwidth]{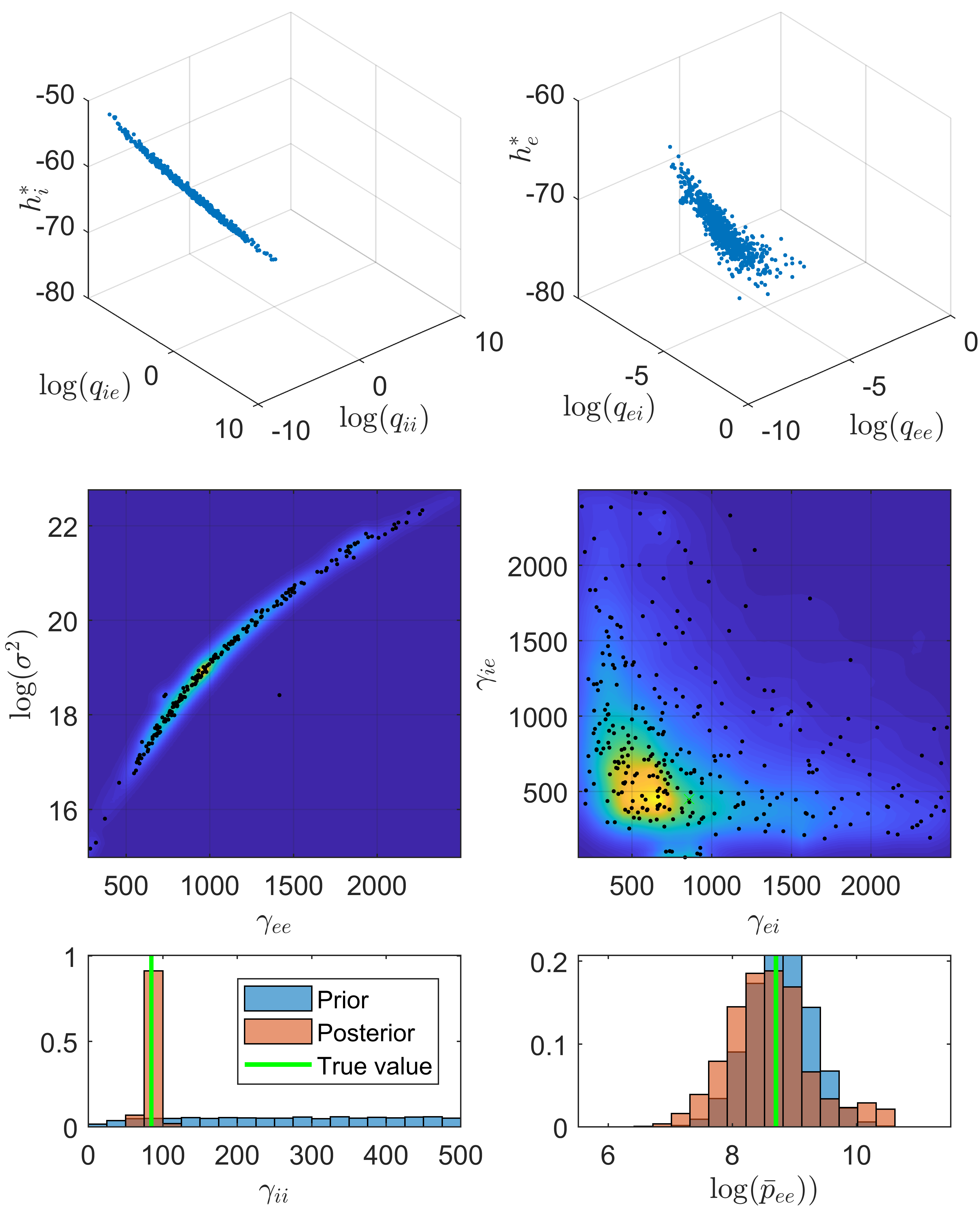}
     \caption{MCMC results with smMALA algorithm for reparameterized
       Neural Population Model with reparameterization.  See Tab.~\ref{tab:npm}
       in the Supplement for the true parameter values,
       and Sec.~\ref{sec:npm_res} for algorithm parameters.}
      \label{fig:langevin_res2}
\end{figure}

\clearpage

Figure \ref{fig:langevin_res2} shows samples obtained from the smMALA
algorithm with the reparameterization.  The algorithm was initialized
from a randomly generated initial parameter set.  After around 2,000
iterations, the sampler finds a region of the parameter space with
approximately the same likelihood as the true parameters.  The
algorithm samples close to the true parameter values for some of the
time, but the range of parameter sets sampled in regions of high
likelihood is relatively wide, indicating weak identifiability and/or
non-identifiability of the model parameters.  The samples generated
from the smMALA algorithm confirm previous results obtained from
sensitivity analysis \citep{Bojak2005}, which found that the model is
more sensitive to the value of $\gamma_{ii}$ than to other parameter
values, see our Figure \ref{fig:langevin_res2}.  More interestingly,
the samples can be used to identify previously unknown relationships
between parameter values in the model.  For example, there is strong
negative correlation between the $h_i$ steady-state value and
$q_{ii}$, which is proportional to the total charge transferred at
synapses between inhibitory neurons, see Figure
\ref{fig:langevin_res2}.
\begin{table}
  \centering
    \begin{tabular}{| l | c | c | c | c |}
    \hline & Acceptance rate & CPU time per iteration & Mean ESS
    & $h$ \\ \hline Original parameterization & 34\% & 5.66s & 36 & 0.3
    \\ Reparameterization & 69\% & 3.18s & 239 & 0.3 \\ Original
    parameterization & 76\% & 5.66s & 23 & 0.1 \\ Reparameterization &
    74\% & 3.18s & 34 & 0.1 \\ \hline
    \end{tabular}
      \caption{Comparison of MCMC with and without model
        reparameterization for smMALA algorithm on NPM.  See Sec.~\ref{sec:npm_res}
        for algorithm parameters.  The parameter $h$
        is the step-size in the smMALA algorithm.}
    \label{tab:comparison2}
\end{table}


\section{Discussion} \label{sec:discussion}

Whilst parameter estimation for ODEs and SDEs is a well-studied
problem in statistics, there has been little work on tailoring
inference techniques to the case where the system operates close to a
stable equilibrium point.  An exception to this is the work contained
in \cite{meerbach2009off}.  Our work is different from this in that we
retain the ability to infer mechanistic parameters of the system, thus
retaining interpretability of model inferences.

This paper exploits the mathematical structure that is present in
quasi-linear dynamics around a stable equilibrium.  We have found that
a reparameterization of the parameter space around an equilibrium
point results in significantly improved efficiency of MCMC algorithms,
and that linearizing the system about the stable equilibrium point
yields a model that approximates the posterior distribution found from
the corresponding nonlinear model.

This approximation can result in a reduction in computation cost of
several orders of magnitude, a welcome boon for any application and a
necessary condition for some to achieve practical feasibility.  For
example, EEG data for a single subject may consist of $50-100$ time
series, each containing 1-2 million data-points. For the data in this
paper, the particle MCMC we ran on a time series of length $200$, took
over a month to run (in R).  Although this may have been speeded up
significantly using the most recent algorithmic and computational
methods in the literature, it is still orders of magnitude slower than
would be required to run on real data. Even the much cheaper EKF is
not feasible for fitting a sophisticated model such as that in Sec.~\ref{sec:npm},
where the dimension of the state space results in
costly matrix inversions at every iteration of the filter. Our
approach avoids these computationally expensive steps. Compared to the
EKF approach, the effect of the linearization about a fixed point
vanishes as the input noise decreases, and the effect of the Whittle
approximation vanishes as the length of the time series
increases. Outside of these situations, our approach tends to
over-estimate posterior uncertainty, but captures the qualitative
dependency between parameters well. Since our approach overestimates
uncertainty, it is well suited to being used to speed up an exact
approach such as importance sampling or delayed acceptance (these
approaches are being compared in \cite{vihola2016importance}).

In addition to the central idea of exploiting the properties of the
stable equilibrium point, we have introduced two additional
innovations that we expect to be more widely useful. Firstly, we have
presented the methodology for using gradient based methods with the
Whittle likelihood for linear SDEs (which may also be used in maximum
likelihood estimation). Secondly, we introduce an easy to calculate
diagnostic for assessing when the Whittle likelihood is likely to be
appropriate.

Whilst the methods in this paper are suitable when the model only has
a single stable equilibrium point, or if the system stays close to
only one specific equilibrium point during observations, they can be
inaccurate when this condition is not satisfied.  We have shown that
it is possible to determine whether the linearization methods are
accurate without having to do a full posterior inference using exact
methods.  The FitzHugh-Nagumo system is often used as an example of a
simple system that exhibits complicated nonlinear behaviour, and for
this system one can construct cases where our proposed linearization
is not appropriate.  For such situations, when particle MCMC is
infeasible, alternative approximate inference techniques are required.
Such methods might build on the work in this paper through the use of
a local Whittle approximation (such as that in \cite{Everitt2013}).

\section*{\hfil Supplementary material \hfil}

\renewcommand{\thesection}{S\arabic{section}}
\renewcommand{\thetable}{S\arabic{table}}
\renewcommand{\thefigure}{S\arabic{figure}}
\renewcommand\theequation{S\arabic{equation}}

\setcounter{section}{0}
\setcounter{table}{0}
\setcounter{figure}{0}
\setcounter{equation}{0}

\section{Introduction}

This supplement contains mathematical detail and derivations, pseudo-code of the MCMC algorithms, as well as discussions of additional results only briefly summarized in the main paper.


\section{Methodology}

\subsection{Model reparameterization}
\label{sec:reparam_supp}

If the dimensions of $\mathbf{F}$ and $\mathbf{\theta}_d$ do not
coincide, e.g., if in our application a background input is added only
to some equations of the system, then the equilibrium calculation can
be split as follows
\begin{align}
\mathbf{F}_1[\mathbf{x}^*, \theta_s]&= 0,
\label{eq:ss2a}\\
\mathbf{F}_2[\mathbf{x}^*, \mathbf{\theta}_s] + \mathbf{\theta}_d &=
0,
\label{eq:ss2b}
\end{align}

\noindent where $\mathbf{F} = (\mathbf{F}_1, \mathbf{F}_2)$, and now
$\mathbf{F}_2$ and $\mathbf{\theta}_d$ have the same dimension, i.e.,
in our application only these equation receive background inputs.  The
previous Eq.~\eqref{eq:ss} then is the special case of $\mathbf{F}_1 =
\emptyset$. Now the model can be reparametrised in terms of
$(\mathbf{\theta}_s, \mathbf{x}_2^*)$, where $\mathbf{x}^* =
(\mathbf{x}_1^*, \mathbf{x}_2^*)$ and $\mathbf{x}_2^*$ has the same
dimension as $\mathbf{\theta}_d$.  Please note that while
$\mathbf{F}_1$ and $\mathbf{x}_1^*$ have the same number of
dimensions, as do $\mathbf{F}_2$ and $\mathbf{x}_2^*$, the
partitioning of state components can differ between $\mathbf{F}$ and
$\mathbf{x}^*$, since we can freely choose which components of
$\mathbf{x}^*$ to trade for the $\mathbf{\theta}_d$.  We can write our
solutions as
\begin{align}
\mathbf{x}_{1}^*&= \mathbf{G}_1[\theta, \mathbf{x}_2^*],
\label{eq:ss2as}\\
\mathbf{\theta}_{d} &= \mathbf{G}_2[\theta_{s}, \mathbf{x}^*] =
\mathbf{G}_2[\theta_{s}, (\mathbf{G}_1[\theta_{s},
    \mathbf{x}_2^*],\mathbf{x}_2^*)] \equiv
\tilde{\mathbf{G}}_2[\theta_{s}, \mathbf{x}_2^*],
\label{eq:ss2bs}
\end{align}

\noindent Note that we assume only that a solution can be obtained
numerically, where $\mathbf{G}_1$ is the solution corresponding to
$\mathbf{F}_1$, and $\mathbf{G}_2$ to $\mathbf{F}_2$,
respectively. The number of unknowns and the number of constraining
equations is the same as in the original equations for the
steady-state: $\mathrm{dim}(\mathbf{x}_1^*) + \mathrm{dim}(\theta_d) =
\mathrm{dim}(\mathbf{x}^*) = \mathrm{dim}(\mathbf{F}) $.
If the solutions in Eqs.~\eqref{eq:ss2as}-\eqref{eq:ss2bs} can be
obtained in closed form by algebraically re-arranging the equations,
then this obviously has practical advantages.
This will be the case for the examples we consider in
the main paper. In this case given $(\theta_s,\mathbf{x}_2^*)$ through an MCMC
update, we can compute $(\mathbf{x}_1^*,\theta_d)$ directly by
substitution into known formulae.

When and why might a reparameterization be effective?
Consider sampling from a complex target distribution with an MCMC
algorithm where the proposal does not take account of the covariance
structure in the target distribution (such as the
Metropolis-within-Gibbs algorithm in supplementary Sec.~\ref{sec:mcmc_alg}). The
steady-state can be a complex function of several elements of
$\theta$, and the likelihood is often very sensitive to
the steady-state.  This leads to strong correlations in the
likelihood.  If the dependencies between elements of
$(\theta_s,\mathbf{x}_2^*)$ are weaker than the dependencies between
elements of $\mathbf{\theta}$, then it will be easier to sample on the
reparametrised space than on the original space.
The reparameterization method can also be useful in MCMC algorithms
that do take account of the covariance structure in the target
distribution (such as the simplified manifold MALA algorithm in
supplementary Sec.~\ref{sec:mcmc_alg}).  If the equilibrium point is a
nonlinear function of the model parameters, then this may cause the
target distribution to have non-constant curvature.  In the
reparameterization, $\mathbf{\theta}_{d}=\tilde{\mathbf{G}}_2[\theta_{s}, \mathbf{x}_2^*]$ will be a nonlinear function. If the likelihood is more sensitive
to $\mathbf{x}_2^*$ than it is to $\mathbf{\theta}_{d}$, then we would
expect sampling on the reparametrised space to be more efficient:
the gain in efficiency is
obtained by reducing nonlinearity in the likelihood function.

Prior distributions may be specified on the original parameter space.
In this case, if we are proposing parameters on the reparametrised
space, we need to evaluate a Jacobian determinant in order to
correctly evaluate the prior density in the MCMC acceptance ratio,
\begin{equation}
p(\theta_s, \mathbf{x}^*_2) = p(\theta_s,
\theta_d)\ |\mathcal{J}_\mathbf{T}|, \ \mathrm{where}\ \mathbf{T}:
(\theta_s, \mathbf{x}_2^*) \rightarrow (\theta_s, \theta_d) =
(\theta_s, \tilde{\mathbf{G}}_2[\theta_{s}, \mathbf{x}_2^*]). \\
\end{equation}
The Jacobian determinant, $|\mathcal{J}_\mathbf{T}|$, can be evaluated
analytically if $\tilde{\mathbf{G}}_2$ is available in closed form.
If not, $|\mathcal{J}_\mathbf{T}|$ can be approximated numerically
using finite differences (yielding a noisy MCMC algorithm
\citep{Alquier2016}, not considered further in this paper).

Examples of applying the reparameterization method, along with further
specific explorations of its effectiveness, can be found in Secs.~\ref{sec:fhn}
and \ref{sec:npm} of the main paper.  However, we want to draw attention
to two further issues that are relatively common in practice.  First,
it is possible that the map $\theta \rightarrow \mathbf{x}^*$ is not
well-defined because there are multiple stable equilibria for one
given set of parameters $\theta$.  In this case the following map is
still well-defined for stable ODEs: $(\theta, \mathbf{x}_0)
\rightarrow \mathbf{x}^*$, where $\mathbf{x}_0$ is the initial
condition of the system.  For SDEs with multiple stable equilibria,
there is always a positive probability that the system will jump
between the basins of attraction of the different stable states.  The
methods presented in this paper are theoretically not applicable in
this setting.  However, in practice the system may remain close to one
specific stable state during the entire interval when the
observations, $y$, are collected.  If so, then it is possible to
construct a map $(\theta, y) \rightarrow \mathbf{x}^*$ that is de
facto well defined. The likelihood for this de facto mono-stability
typically increases with a decrease of the strength of the noise
driving the system.  Second, the proposal densities
$q[(\theta_s,\theta_d)\rightarrow(\mathbf{x}^*_1,\mathbf{x}^*_2)]$ and
$\tilde{q}[(\theta_{s},\mathbf{x}_2^*)\rightarrow(\mathbf{x}^*_1,\theta_d)]$
could have support on different regions of
$\mathcal{R}^{d_{\theta}+d_x}$. For example, prior knowledge may
dictate that for $\theta_d < 0$ one has $q = 0$ but this is not
explicit in the reparametrised $\tilde{q}\neq 0$.  If there is such
strong prior information then more work needs to be done to ensure
that $q$ and $\tilde{q}$ have the same support.  For example, it may
be that $\theta_d > 0$ implies $\mathbf{x}^*_2 > \mathbf{x}_0$.  If
so, then an appropriate
$\tilde{q}[(\theta_{s},\mathbf{x}_2^*)\rightarrow(\mathbf{x}^*_1,\theta_d)]$
would be a normal distribution on $\log[ \mathbf{x}^*_2 - \mathbf{x}_0
]$.

\subsection{The spectral density of stable SDEs}
\label{sec:spec_dens}

\subsubsection{Linearization and the spectral density}
\label{sec:spec_dens_lin}
In Sec.~\ref{sec:Whittle} we describe several different methods for
approximating the likelihood in stable SDEs.  In this section, we introduce two
key tools that are needed for the faster likelihood calculations.
The first tool is linearization.  To linearize the model, we apply a Taylor
approximation to the right-hand side of Eq.~\eqref{eq:nonlinear_de}
\begin{equation} \label{eq:taylor}
 \mathbf{F}[\mathbf{x}; \theta] \approx \mathbf{F}[\mathbf{x}^*;\theta] + \mathcal{J}(\mathbf{x}^*; \theta) (\mathbf{x} - \mathbf{x}^*),
\end{equation}
\noindent where $\mathcal{J}(\mathbf{x}^*, \theta)$ is the Jacobian evaluated at
$\mathbf{x}^*$ with parameters $\theta$.
The accuracy of linearization depends on the size of the higher order
derivatives of $\mathcal{J}$, and on the size of $\delta\mathbf{x}\equiv\mathbf{x} - \mathbf{x}^*$.  We are
seeking to apply linearization around a stable equilibrium $\mathbf{F}[\mathbf{x}^*;\theta]=0$.
The size of $\delta\mathbf{x}$ will then depend on the size of $\mathbf{p}(t; \theta)$, which is the term that
pushes the system away from its equilibrium:
\begin{equation} \label{eq:linear_de}
\frac{d}{d t} \delta\mathbf{x}(t) = \mathcal{J}(\mathbf{x}^*; \theta)\ \delta\mathbf{x}(t)+ \mathbf{p}(t;
\theta),
\end{equation}
The effect of this approximation on inference was explored in
Sec.~\ref{sec:particle_v_marginal} of the main paper.

The second tool is a spectral density calculation.  This is useful for
performing inference in the frequency domain, e.g., evaluation of the Whittle
likelihood.  If $x_j(t)$ is the $j\/$-th component of the multivariate stochastic process
$\mathbf{x}(t)$, then the (power) spectral density of $x_j(t)$ is
\begin{equation}
\label{eq:psd}
S_{xx}^j(\nu) = \frac{1}{T}\left|X_j^T(\nu)\right|^2,
\end{equation}
\noindent i.e., the spectral density is proportional to the squared modulus of $X_j^T(\nu)$,
where $X^T_j(\nu)$ denotes the Fourier transform of $x_j^T(t)$. Here $x_j^T(t)$ is $x_j(t)$
with support on the interval $\left[-\frac{T}{2},\frac{T}{2}\right]$, i.e., with $x_j^T(t)=0$ outside
of this interval. In principle Eq.~\eqref{eq:psd} is to be understood in
a $\lim_{T\rightarrow\infty}$ sense, but in practice $T$ corresponds to the actual length
of the data and one obtains an estimate.
To simplify the notation we will drop the component index $j$ and the interval superscript $T$ in the following.
The Fourier transform is defined only up
to choices in constant factors and naming conventions \citep{fouriertrans}:
\begin{equation}
x(\beta)=\sqrt{\frac{|b|}{(2\pi)^{1+a}}}\int_{-\infty}^\infty d\alpha\, X(\alpha)e^{-ib\alpha\beta}
\quad\mathrm{and}\quad
X(\alpha)=\sqrt{\frac{|b|}{(2\pi)^{1-a}}}\int_{-\infty}^\infty dt\, x(\beta)e^{ib\alpha\beta}.
\end{equation}
In Eq.~(\ref{eq:psd}) we have implicitly used $(a,b,\alpha,\beta)=(0,-2\pi,\nu,t)$, which is common in
applications. For this choice the Fourier variable $\alpha$ corresponds to the ordinary frequency $\nu$.

Assuming a weakly stationary process mean-zero process, and here setting $\beta=\tau$ as is convention
for the lag variable, the Wiener-Khinchin theorem then can be written as
\begin{equation}
\label{wktheorem}
\gamma_{xx}(\tau) = \mathrm{E}\left[x(t+\tau)x(t)\right] = \int_{-\infty}^\infty d\nu\ S_{xx}(\nu) e^{i2\pi \nu\tau},
\end{equation}
i.e., the spectral density is simply the Fourier transform of the autocovariance.
Typically one assumes that the difference between the (discretely sampled) process and the observations is a stochastic process,
such as white noise. When two processes $x(t)$ and $z(t)$ are independent,
then the autocovariance of $y(t)=x(t)+z(t)$ is simply $\gamma_{yy}(\tau)=\gamma_{xx}(\tau) + \gamma_{zz}(\tau)$.
It follows from Eq.~\eqref{wktheorem} that then also $S_{yy}(\nu)=S_{xx}(\nu) + S_{zz}(\nu)$.
If $z(t)$ is zero-mean white noise, then $\gamma_{zz}(\tau)=\sigma^2\delta(\tau)$ with
variance $\sigma^2$ and Dirac $\delta(t)$, and consequently the measured spectral density will have
a ``noise floor'' $S_{yy}(\nu)=S_{xx}(\nu)+\sigma^2$. The connection to the discrete (time) Fourier transform required
for analysing experimental data is straightforward, using $\Delta t=T/n$ and $\Delta\nu=1/T$:
\begin{align}
x\left(t=t_l\equiv l\cdot\Delta t\right) &\quad\longrightarrow\quad x_l=\frac{1}{\sqrt{n}}\sum_k X_k e^{i2\pi kl/n}, \\
X\left(\nu=\nu_k\equiv k\cdot\Delta\nu\right)/\Delta t&\quad\longrightarrow\quad X_k=\frac{1}{\sqrt{n}}\sum_l x_l e^{-i2\pi kl/n},\\
S_{xx}\left(\nu=\nu_k\equiv k\cdot\Delta\nu\right)/\Delta t&\quad\longrightarrow\quad S_k=|X_k|^2,
\end{align}
where $k,l=-\frac{n}{2}+1,..,\frac{n}{2}$, or $k,l=0,...,n-1$ with the explicit
$n\/$-periodicity in the discrete case.

The Fourier transform of a system of linear SDEs can be evaluated as follows,
where we will be using the Fourier choice common in mathematics
$(a,b,\alpha,\beta)=(1,-1,\omega,t)$ to avoid extraneous ``$2\pi$'' factors.
Any system of linear differential equations can be written in the form
\begin{equation} \label{eq:de}
\frac{d}{dt}\mathbf{x}(t) = \mathcal{A}\, \mathbf{x}(t) + \mathbf{p}(t),
\end{equation}
cf.\ Eq.~\eqref{eq:linear_de}. Using integration by parts, the Fourier transform of $d\mathbf{x} / dt$ is
\begin{align}
\int_{-\infty}^{\infty} dt\ \left(\frac{d}{dt}\mathbf{x}(t)\right) e^{ -i \omega t }&=
\bigg[\mathbf{x}(t)  e^{ -i \omega t } \bigg]_{t=-\infty}^{t=\infty} -
\int_{-\infty}^{\infty} dt\ \mathbf{x}(t) \left( \frac{d}{dt} e^{ -i \omega t }\right) \nonumber\\
&= i\omega\mathbf{X}(\omega).
\end{align}
Applying the Fourier transform also to the other side of Eq.~\eqref{eq:de} we obtain
\begin{equation} \label{eq:de_ft}
\left(i\omega\mathbb{1}-\mathcal{A}\right)\mathbf{X}(\omega) = \mathbf{P}(\omega).
\end{equation}
For a given value of $\omega$, Eq.~\eqref{eq:de_ft} is a linear system of
equations, which can be re-arranged to obtain $\mathbf{X}(\omega)$ as a
function of $\mathbf{P}(\omega)$.  The matrix $\mathcal{T}(\omega)$
that maps $\mathbf{P}(\omega)$ to
$\mathbf{X}(\omega)$ is called the transfer function
\begin{equation}
\mathbf{X}(\omega) = \mathcal{T}(\omega)\ \mathbf{P}(\omega).
\end{equation}
An expression for $\mathcal{T}(\omega)$ can be obtained using the
eigen-decomposition of $\mathcal{A}$ \citep{Bojak2005}:
\begin{equation} \label{eq:transfer}
\mathcal{T}(\omega) = \mathcal{R}\ \mathrm{diag}\bigg[ \frac{1}{e_k (i\omega - \lambda_k)} \bigg]\ \mathcal{L},
\end{equation}
\noindent where the $j\/$-th column of $\mathcal{R}$ is the $j\/$-th
right-eigenvector of $\mathcal{A}$, the $i\/$-th row of $\mathcal{L}$ is the $i\/$-th left-eigenvector,
$\lambda_1,\ldots,\lambda_k,\ldots,\lambda_d$ are its eigenvalues, and $e_1,\ldots,e_k,\ldots,e_d$ are
norms of the eigenvectors obtained from $\mathcal{L}\mathcal{R}=\mathrm{diag}[e_k]$.

This derivation provides us with an important observation: the computational complexity of evaluating the spectral density can
be reduced if only some of the elements of the transfer function matrix are needed. In the examples in this paper only one
element of $\mathbf{x}(t)$ is observed, and only one element of the input
$\mathbf{p}(t)$ is assumed to be non-zero. This means that only a single
element of the transfer function matrix needs to be calculated.
Thus, by exploiting sparsity in both the assumed input to the system and its measured output,
the complexity of evaluating the spectral density is reduced to $O(N\, d)$, with $N$ the number of points
where the spectral density is being evaluated and $d$ the dimension of the state-space. This implies
significant computational savings compared to the worst case $O(N\, d^3)$ where all elements of
$\mathbf{x}(t)$ and $\mathbf{p}(t)$ are needed. Such sparsity in input-output components
is quite common in scientific applications, since one one hand often experimentally only
one or a small number of state variables of the system are accessible, and on the other hand
experiments are often designed to disentangle inputs (e.g., a task provides
highly specific excitation or observations are made when a certain inputs are known to be dominant).

Other straightforward ways of improving the efficiency of such spectral density calculations are
\begin{itemize}
\item reducing the frequency resolution, e.g., by decimation or window-averaging;
\item setting the spectral density to zero above a known frequency threshold of the system; and
\item parallelizing the computation of the $d^2$ entries of $\mathcal{T}(\omega)$.
\end{itemize}
Use of the first two of techniques is problem-dependent, and care needs to be taken to ensure that the bias introduced is negligible.
The results here can be easily extended to calculate cross-spectral densities.  This is useful for problems where multiple components of the process are observed.

\subsubsection{Derivatives of the spectral density}

Some parameter estimation algorithms, such as Metropolis Adjusted Langevin
Algorithm (MALA) require the likelihood to be differentiated.  In order to apply
such algorithms to likelihoods based on the spectral density, it is necessary to
differentiate the spectral density.  Derivatives can be approximated using
finite differences.  However, these can be time-consuming to compute, and inaccurate.
Furthermore, the
step-size in the finite difference approximation can only be decreased to a certain extent
before rounding errors occur due to the limited machine precision.
Here we derive analytic derivatives for the spectral density.  The results are only valid when all eigenvalues of the system are distinct.  In the problems we have looked at this condition is satisified for most of the parameter space, but not all of it.  It is always possible to fall back on finite differences, if parameter sets with repeated eigenvalues are sampled.

The derivatives of the spectral density can be used to evaluate analytic derivatives of the Whittle likelihood for stable SDEs as follows, where we indicate the derivative by the parameters $\theta$ as subscript. Note that the parameters themselves are real, not complex.
\begin{gather}
T S_{xx,\theta}(\omega) = \frac{\partial}{\partial \theta} \left|\mathbf{X}(\omega)\right|^2
=\left\{\frac{\partial}{\partial\theta}\left[\Re^2\mathcal{T}(\omega)+\Im^2\mathcal{T}(\omega)\right]\right\}\left|\mathbf{P}(\omega)\right|^2+
\left|\mathcal{T}(\omega)\right|^2\left\{\frac{\partial}{\partial\theta}\left[\Re^2\mathbf{P}(\omega)+\Im^2\mathbf{P}(\omega)\right]\right\}\nonumber\\
=2\left\{\left[\Re\mathcal{T}(\omega) \Re\mathcal{T}_\theta(\omega)+\Im\mathcal{T}(\omega)\Im\mathcal{T}_\theta(\omega)\right]\left|\mathbf{P}(\omega)\right|^2+
\left|\mathcal{T}(\omega)\right|^2\left[\Re\mathbf{P}(\omega)\Re\mathbf{P}_\theta(\omega)+\Im\mathbf{P}(\omega)\Im\mathbf{P}_\theta(\omega)\right]\right\}
\end{gather}
To evaluate the derivatives of $\mathcal{T}(\omega)$, we need the derivatives
of the eigenvalues and eigenvectors \citep{Giles1948}. Differentiating both sides of
$\mathcal{J}\mathbf{r} = \lambda \mathbf{r}$ gives us,
\begin{equation} \label{eq:eig_diff}
\mathcal{J}_\theta \mathbf{r} + \mathcal{J} \mathbf{r}_\theta = \lambda_\theta \mathbf{r} + \lambda \mathbf{r}_\theta,
\end{equation}
\noindent where $\mathcal{J}_\theta$ means each element of the Jacobian matrix $\mathcal{J}$ is differentiated with respect to $\theta$, and likewise for
the right eigenvector $\mathbf{r}$ and $\mathbf{r}_\theta$.
We require that right eigenvectors have unit length, i.e., $\mathbf{r}^T\mathbf{r} = 1$.
Differentiating  this constraint with respect to $\theta$ yields
\begin{equation} \label{eq:eig_con}
\mathbf{r}^T \mathbf{r}_\theta = 0.
\end{equation}
Combining Eqs.~\eqref{eq:eig_diff} and \eqref{eq:eig_con}, and re-arranging, we obtain,
\begin{equation}
 \left[
 \arraycolsep=2pt\def\arraystretch{2.5}
\begin{array}{c|c}
\mathcal{J} - \lambda \mathbb{1} & -\mathbf{r} \\
\hline
\mathbf{r}^T & 0  \\
\end{array}
\right]
\left[
 \arraycolsep=2pt\def\arraystretch{2.5}
\begin{array}{c}
\mathbf{r}_\theta  \\
\hline
\lambda_\theta  \\
\end{array}
\right]
=
\left[
 \arraycolsep=2pt\def\arraystretch{2.5}
\begin{array}{c}
- \mathcal{J}_\theta\ \mathbf{r}  \\
\hline
0  \\
\end{array}
\right].
\end{equation}

\noindent This is a $(d+1)\times(d+1)$ linear system, which we can solve for
$(\mathbf{u}_\theta,\lambda_\theta)$.  Furthermore, if we want to calculate
derivatives with respect to $p$ different parameters, we can calculate the
inverse of the matrix on the left-hand side, which is an $\mathcal{O}(d^3)$ operation, and use
this matrix to calculate the $p$ derivatives.  Overall the computational
complexity is then $\mathcal{O}(d^3 + p d^2)$.
The cubic complexity here is manageable in the regime we are interested in, i.e., when $p,d \approx 10-50$.
In general the rate-limiting steps in our calculations are those that involve a factor of $n$, as $n > 10^4$.

When $p$ and $d$ are both on the order of $10-100$ or greater, this means we get
approximately a factor of $d$ saving compared to the finite difference approach,
which evaluates the eigenvalues and eigenvectors separately for each of the $p$
parameters. A similar calculation can be done for left eigenvectors,
$\mathbf{l}^T\mathcal{J} = \lambda \mathbf{l}^T$, and the computational
saving is the same:
\begin{equation}
\mathbf{l}^T_\theta\left(\mathcal{J} - \lambda \mathbb{1}\right)=-\mathbf{l}^T\left(\mathcal{J}_\theta - \lambda_\theta \mathbb{1}\right).
\end{equation}
Note that we assume here that $\lambda_\theta$ has already been computed in the
right eigenvector case. Furthermore, we require the cross-norm $\mathbf{l}^T\mathbf{r} = 1$, i.e.,
$\mathcal{L}\mathcal{R}=\mathrm{diag}[e_k]=\mathbb{1}$. This removes any further parameter dependence
from $\mathcal{T}(\omega)$.
For second derivatives, we can simply differentiate Eqs.~\eqref{eq:eig_diff}
and \eqref{eq:eig_con} again, to obtain another $(d+1)\times(d+1)$
linear system.  This time the second derivatives of the eigenvectors and
eigenvalues are the unknowns.  In this case the overall computational complexity
for calculating the second derivatives with respect to $p$ different parameters
is $\mathcal{O}(d^3 + p^2 d^2)$, compared to $\mathcal{O}(p^2 d^3)$ for the finite difference
calculation. Once the eigenvector and eigenvalue derivatives have been obtained it is
straightforward to differentiate $\mathcal{T}(\omega) = \mathcal{R}\ \mathrm{diag}[ 1 / (i\omega - \lambda_k)]\ \mathcal{L}$.
In the  worst-case scenario when all elements of $\mathcal{T}(\omega)$ are
needed, this has a computational complexity of $\mathcal{O}(p\, n\, d^3)$ for first
derivatives.  In the best-case scenario where only one element is needed,
it is $\mathcal{O}(p\, n\, d)$. Differentiating with respect to the parameters of the input process,
$\mathbf{P}(\omega)$, is straight-forward because the parameters values in
$\mathbf{P}(\omega)$ do not affect the eigen-decomposition of $\mathcal{J}$.

\subsection{Marginal likelihoods for indirectly observed SDEs}
\label{sec:like_supp}
\subsubsection{Particle filter}
\label{sec:pf_supp}
If we discretize an SDE, and assume that observations are a sample
path plus some observation noise, we obtain a state-space model.  The
solutions to SDEs and state-space models are samples from a
probability distribution over an appropriate space.  To simplify the
notation we write the distribution explicitly only in the
finite-dimensional state-space case.  The marginal likelihood of
$y_{1:n}$, conditional on $\theta$ is,
\begin{equation}
p_{\theta}(y_{1:n}) = p_{\theta}(y_1) \prod_{i=2}^n p_{\theta}(y_i |
y_{1:i-1} )
\end{equation}

\noindent where
\begin{equation} \label{eq:marg_int}
p_{\theta}(y_i | y_{1:i-1} ) = \int g_{\theta}(y_i|x_i)
f_{\theta}(x_i|x_{i-1}) p_{\theta}(x_{1:i-1}|y_{1:i-1})\, dx_{1:n}
\end{equation}

\noindent Particle filters combine importance sampling and resampling
to obtain Monte Carlo estimates of the integral in
Eq.~\eqref{eq:marg_int} for $i$ from $2$ to $n$.  In the simplest form
of the particle filter, the importance sampling proposal is
$f_{\theta}(x_i|x_{i-1}) \hat{p}_{\theta}(x_{1:i-1}|y_{1:i-1})$, and
the importance weights are $g_{\theta}(y_i|x_i)$.  The resampling uses
the normalized weights as probabilities of a discrete distribution.
The distribution $\hat{p}_{\theta}(x_{1:i-1}|y_{1:i-1})$ is a sum of
equally weighted Dirac delta-functions centred on the resampled
particles. If we use $N$ particles to estimate the integrals in
Eq.~\eqref{eq:marg_int}, the overall cost of the estimate is $O(N n
d)$, where $d$ is the dimension of the state-space.  Depending on the
problem and the variant of particle filter used, $N$ may be somewhere
in the range 100-10,000.  Outside of special cases where we may
simulate exactly from the SDE dynamics, this approach yields a biased
approximation to the likelihood due to discretization error. If we
wish for a better approximation to the SDE, we may sample paths on a
finer grid than the observations, which increases the computational
cost of the particle filter.

\subsubsection{Kalman filter}

If the SDE or state-space model is linear then
$f_{\theta}(x_i|x_{i-1})$ is a Gaussian distribution. If we further
assume that the likelihood of observations $y_{1:n}$ conditional on
$x_{1:n}$ is a product of Gaussians, then the probability
distributions inside the integral of Eq.~\eqref{eq:marg_int} are all
Gaussian, and the Kalman filter can be used to evaluate the marginal
likelihood analytically.  This has a computational cost of $O(n
d^3)$. Again the cost is higher if we sample paths on a finer grid
than the observations. We note that in the ODE case, where there is no
uncertainty in the $x_{1:n}$ conditional on $\theta$, the likelihood
simplifies to the evaluation of a multivariate Gaussian, with
computational cost $O(n d)$:
\begin{equation} \label{eq:dl}
p_{\theta}(y_{1:n}) = \prod_{i=1}^n p\left[y_i | x_{\theta}(t_i)
  \right] = \prod_{i=1}^n \mathrm{N}\left[x_{\theta}(t_i),
  \sigma_o^2\right].
\end{equation}

The standard approach to linearizing a non-linear dynamic system is
the \emph{extended Kalman filter} (EKF), where a linearization of the
dynamics is performed at each iteration of the filter, setting the
transition matrix to be the Jacobian evaluated at the current state
estimate. There are some applications where this approximation is
known to be inadequate, leading to the divergence of the
filter. However, in practice this approximation is often suitable, and
the method is widely used.  In this paper we propose to use an
approximation that is less widely applicable, but which we show can
lead to very computationally efficient algorithms. Namely, we propose
to use a single linearization of the dynamics, about the stable
equilibrium point of the model (as described in the supplementary
Sec.~\ref{sec:spec_dens}). This yields a single transition matrix
to be used at every iteration of the Kalman filter.  Although this
approximate model is usually less accurate than the EKF, as alluded to
in Sec.~\ref{sec:de_defs}, the additional approximation can be
small (as has been exploited in computational neuroscience, e.g., in
\cite{Moran2009, Pinotsis2012, Abeysuriya2015}).

\subsubsection{Whittle likelihood}
\label{sec:whittle_def}

In the remainder of this section we briefly recall the derivation of the
Whittle likelihood, following \cite{RobertH.Shumway2011}.  We will
refer to this in Sec.~\ref{sec:kalman_v_whittle} of the Supplement,
where we study the accuracy of this approximation.  We assume that the
process has zero mean, but it is not difficult to generalize the
Whittle likelihood to the non-zero case. As defined above, the
periodogram of a time series is the squared modulus of its Discrete
Fourier Tranform (DFT)
\begin{align} \label{eq:wl}
S_k=|X_k|^2 &= \left|\sum_l x_l e^{-i2\pi
  kl/n}\right|^2\nonumber\\ &=\left[\frac{1}{\sqrt{n}}\sum_l x_l
  \cos\frac{2\pi kl}{n}\right]^2+\left[\frac{1}{\sqrt{n}}\sum_l x_l
  \sin\frac{2\pi kl}{n}\right]^2\nonumber\\ &= X_k^{c\;2} +
X_k^{s\;2}.
\end{align}

\noindent with the the discrete cosine and sine transforms. If the
time series $x_l$ is a normal random variable with mean zero, then as
linear combinations also $X_k^c$ and $X_k^s$ will be jointly normal
with mean zero.  Using $f(\nu)\equiv S_{xx}(\nu)/\Delta t$ and
$\nu_k=k\cdot\Delta \nu=k/(n\Delta t)$, the variances and covariances
are as follows \citep{RobertH.Shumway2011}:
\begin{align}
\mathrm{Cov}[X_k^c, X_l^c] &= \frac{f(\nu_k)}{2} \delta_{kl} +
\epsilon_n \label{eq:w2} \\ \mathrm{Cov}[X_k^s, X_l^s] &=
\frac{f(\nu_k)}{2} \delta_{kl} + \epsilon_n\\ \mathrm{Cov}[X_k^c,
  X_l^s] &= \epsilon_n, \label{eq:w3}
\end{align}

\noindent with Kronecker $\delta_{kl}$. Here $|\epsilon_n| < \phi /
n$, and
\begin{equation} \label{eq:supp_phi}
\phi = \sum_{h=-\infty}^{\infty} |h|\left|\gamma_{xx}(h\cdot\Delta
t)\right|,
\end{equation}

\noindent where $\gamma_{xx}(\tau)$ is the autocovariance function
defined in Sec.~\ref{sec:spec_dens_lin} of the supplementary
material.

The Whittle likelihood is obtained by dropping the $\epsilon_n$ terms.
This eliminates the dependence between $X_k^c$ and $X_k^s$ as well as
the bias in the remaining terms.  The sum of the squares of two
independent standard normal random variables is a $\chi^2$
distribution with two degrees of freedom, hence
\begin{equation} \label{eq:supp_whittle}
p_{\theta}(x_0,\ldots,x_{n-1}) = p_{\theta}(S_0,\ldots,S_{n-1})
\approx \prod_{k=1}^{n/2-1} \frac{1}{f(\nu_k;\theta)}
\exp\left[-\frac{S_k}{f(\nu_k;\theta)}\right],
\end{equation}
\noindent where we now have made explicit the dependence on model
parameters $\theta$ and, as is common, neglected the mean and Nyquist
spectral edges $k=0,n/2$.  Note that the terms
$S_{n/2},\ldots,S_{n-1}$ do not appear in the product since they are
are fully determined by $S_{1},\ldots,S_{n/2 - 1}$ for real signals.
The equation $|\epsilon_n| < \phi / n$ tells us that the error in the
approximation decreases linearly with $n$, and also that it depends on
the decay rate of the autocovariance function.  The effect of the
error on inference, and how this varies with $\phi$ and $n$ is
analysed in Sec.~\ref{sec:kalman_v_whittle} of the Supplement.
It is straightforward to extend the derivation above to obtain an
Whittle likelihood for observations
$p_{\theta}(y_0,\ldots,y_{n-1})$, if the system state
$\mathbf{x}_i$ is indirectly observed according to Eq.~\eqref{eq:model2}
and independent from the ``observation noise'' $\mathbf{n}_i$.

\subsection{MCMC algorithms}
\label{sec:mcmc_alg}

\begin{algorithm}[ht]
\KwIn{Data, $y$; Initial value for $\theta = (\theta_1,\ldots,\theta_p)$;
\newline
Likelihood $l(y|\theta)$; Prior distribution $p(\theta)$.}
\Parameter{Proposal variances, $\sigma^2_1$,\ldots,$\sigma^2_p$; Number of
iterations, $I$}
\For{$i = 2,\ldots,I$ }{
\For{$j = 1,\ldots,p $}{
    Propose $\theta^*_j \sim N(\theta_j, \sigma^2_j)$\;
    Evaluate acceptance ratio, $\alpha(\theta, \theta^*) =
        \mathrm{min}\left[1, \dfrac{l(y|\theta^*) p(\theta^*) }{l(y|\theta)
p(\theta) } \right]$\;
    Draw $u \sim \mathrm{Uniform}(0,1)$\;
    \lIf{$u < \alpha(\theta, \theta^*)$}{Set $\theta = \theta^*$ }
}
}
\caption{Metropolis-within-Gibbs (MwG)}\label{alg:mwg}
\end{algorithm}

\begin{algorithm}[ht]
\KwIn{Data, $y$; Initial value for $\theta = (\theta_1,\ldots,\theta_p)$;
\newline
Likelihood $l(y|\theta)$; Prior distribution $p(\theta)$.}
\Parameter{Step size, $h$; Number of iterations, $I$}
\For{$i = 2,\ldots,I$ }{
    Evaluate gradient and Hessian of the unnormalized log posterior,\newline
    \Indp $g = \nabla \bigg[\log[ \ l(y|\theta)\ p(\theta) \ ] \bigg]$ and $H_{\theta}$\;
    Set $C = h^2 H_{\theta}^{-1}$, and $m = \theta + \frac{1}{2} C\, g$\;
    Propose $\theta^* \sim q(\theta^*|\theta)=N(m, C)$\;
    Evaluate acceptance ratio, $\alpha(\theta, \theta^*) =
        \mathrm{min}\left[1, \dfrac{l(y|\theta^*) p(\theta^*) q(\theta|\theta^*)
}{l(y|\theta) p(\theta) q(\theta^*|\theta)} \right]$\;
    Draw $u \sim \mathrm{Uniform}(0,1)$\;
    \lIf{$u < \alpha(\theta, \theta^*)$}{Set $\theta = \theta^*$ }
}
\caption{Simplified manifold Metropolis Adjusted Langevin Algorithm
(smMALA)}\label{alg:mmala}
\end{algorithm}

In this paper we use two standard MCMC algorithms: Metropolis-within-Gibbs (MwG), and the simplified manifold Metropolis Adjusted Langevin
Algorithm (smMALA) \citep{Girolami2011a}. Pseudo-code for these are provided as Algorithms \ref{alg:mwg}
and \ref{alg:mmala}.  The MwG algorithm is easier to implement.  The smMALA gives
a higher Effective Sample Size (ESS) because the proposal uses the local gradient and
Hessian of the posterior.  The smMALA requires more computational resources per
iteration because of the gradient and Hessian calculations.  If the posterior
has a relatively simple geometry, the cost of the gradient and Hessian
calculations may outweigh the benefits of a better proposal.
An accessible presentation of the smMALA algorithm along with application to a brain imaging problem can be found in \cite{Penny2016}.
In all cases the algorithm can be implemented with the
original model parametrisation or with the steady-state reparametrisation
outlined in Sec.~\ref{sec:reparam}. In Secs.~\ref{sec:fhn} and \ref{sec:npm}, we will demonstrate how using the steady-state reparametrisation can lead to more efficient sampling of the parameter space.
When using the Langevin method, we calculate the gradient of the Whittle likelihood using the analytic derivatives of the spectral density derived in Sec.~\ref{sec:spec_dens}.
In later sections, we use the term marginal MCMC as a label for MCMC algorithms where the likelihood $l(y|\theta)$ is a marginal likelihood that is calculated analytically, e.g., using a Kalman filter.

\section{Analysis using the FitzHugh-Nagumo equations}

\begin{figure}[htbp]
\centering
  \includegraphics[width=\textwidth]{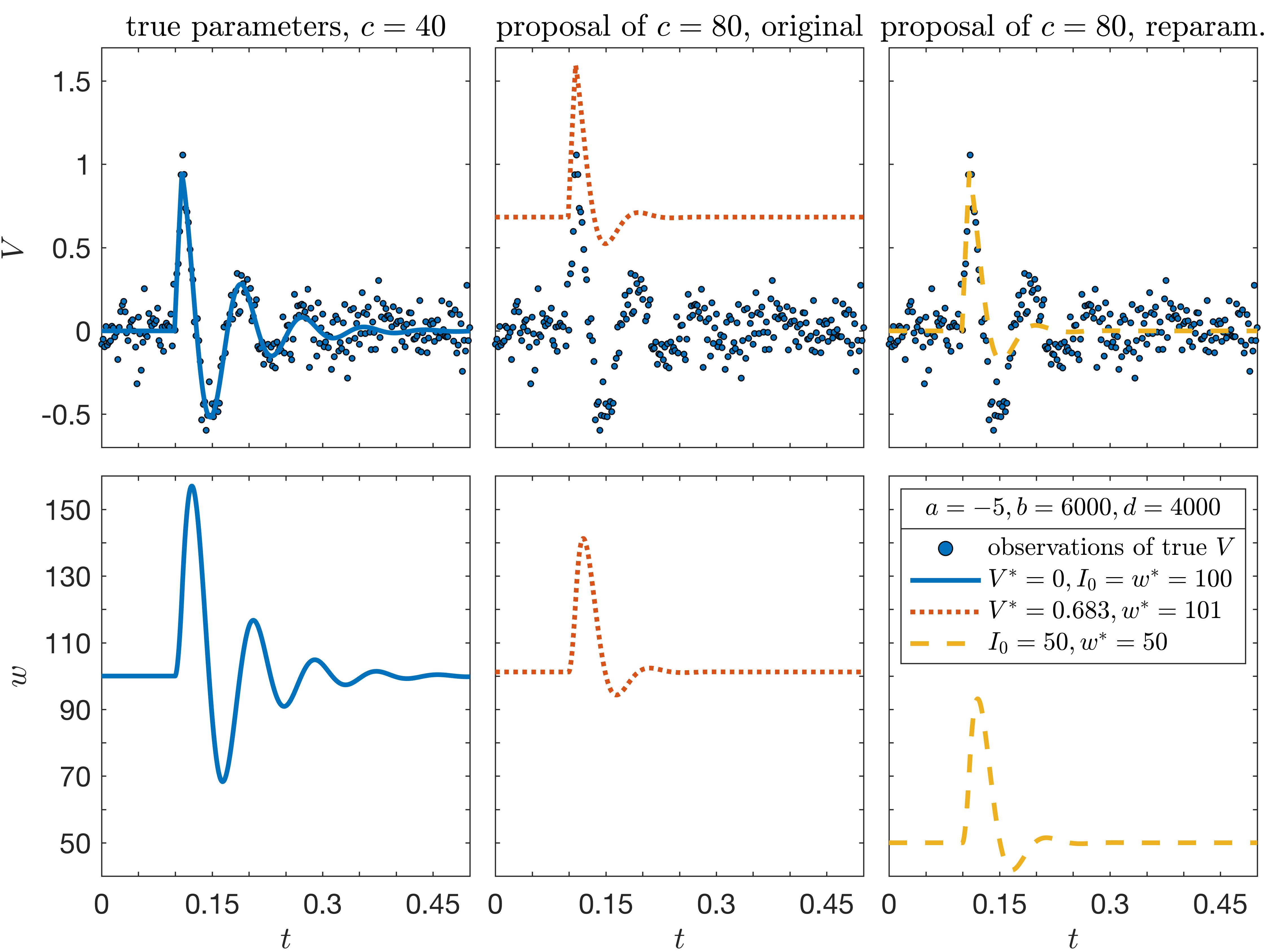}
\caption{Reparameterization example. \textit{Left column}: Solutions (blue solid lines)
for $V$ (top) and $w$ (bottom) for $a=-5$, $b=6000$, $d=100$, $c=40$, $I_0=100$.
Noisy observations of this $V$ (blue dots) are shown
here and in the other columns. \textit{Middle column}: A proposal with $c=80$ (other
parameters identical) changes $(V^*,w^*)$. The
solution (orange dotted line) deviates strongly from the observations. \textit{Right column}:
The same proposal in the reparametrised scheme (golden dashed line) changes $(I_0,w^*)$. Since $V^*=0$ remains unchanged, deviations remain limited.}
\label{fig:fhn1}       
\end{figure}

\subsection{Results for deterministic FitzHugh Nagumo equations}
\label{sec:det_fhn}

\begin{figure}[htbp]
\centering
  \includegraphics[width=\textwidth]{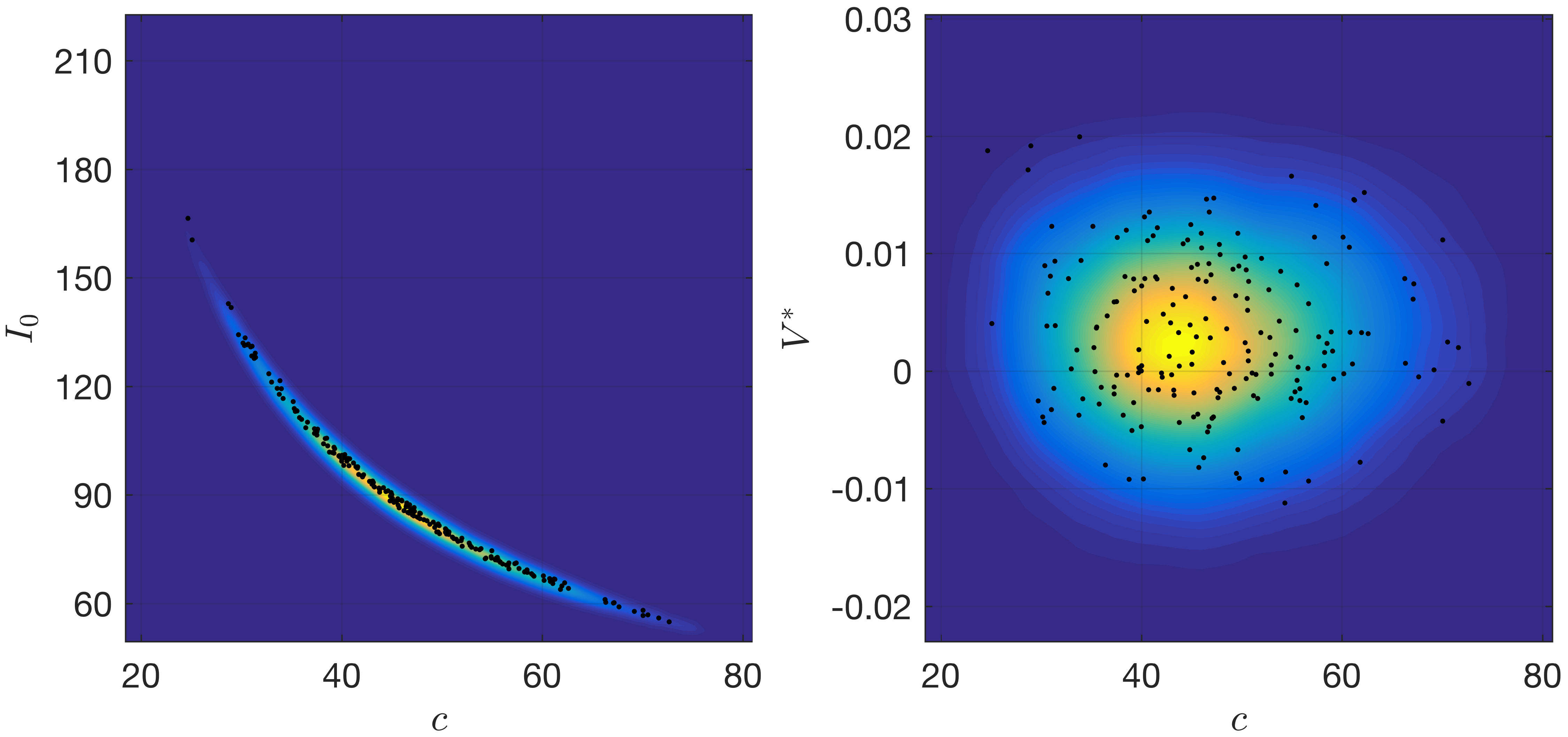}
%
\caption{MCMC results for deterministic FHN.  The posterior correlation between $c$ and $V^*$
(\textit{right panel}, reparametrised) is much weaker than the correlation between $c$ and $I_0$
(\textit{left panel}, original). The MCMC chain was started at the true parameter set
($a=-5$, $b=6000$, $c=40$, $I_0=100$) and run for 100,000 iterations.
Every 500th sample (black dots) and a smoothed histogram (background colour map)
are shown. Quantiles (0.025, 0.5, 0.975) of posteriors for parameters not
shown were $a$: (-11, 4.11, 15.4)
and $b$: (4980, 5400, 6250).}
\label{fig:fhn2}       
\end{figure}

Here we set $P(t)=0$ throughout, but $I(t) = 100$ for the time interval $1 < t < 1.1$ and $I(t) = 0$ otherwise.
The initial conditions are set to the stable equilibrium.  This is a function of the model parameters, so the initial conditions change when the parameters are updated.  It would not be difficult to extend the methods here to the case when the model parameters and the initial conditions are treated as separate parameters of the inference problem.  The numerical
solution and parameter values used are shown in Fig.~\ref{fig:fhn1},
along with numerical solutions at two other parameter sets. We make
the following assumptions about the inference problem: (i) we can only observe $V$ with observation noise, and do not
have any observations for $w$, (ii) the value of $d$ is known, (iii) The prior distributions for $a$, $b$, $c$, $I_0$ are uniform over a large (unspecified) range.

Scatterplots for selected pairs of parameters and table summaries of the posterior distribution
(obtained using the reparametrised algorithm) are shown in Figure
\ref{fig:fhn2}.

Figure \ref{fig:fhn1} illustrates how the likelihood of the data is sensitive to the steady-state value $V^*$.  This induces strong correlations in the posterior distribution between the original model parameters, $c$ and $I_0$, seen in Figure \ref{fig:fhn2}.  Reparameterizing the likelihood using the steady-state value results in a likelihood function with weaker correlations (Figure \ref{fig:fhn2}).  This means that that the MwG algorithm should be able to sample the parameter space more efficiently.
A Metropolis-within-Gibbs MCMC algorithm (Algorithm \ref{alg:mwg}) was run with and without model
reparameterization.  In both cases, the parameters are updated using a Gaussian distribution centred on the
current parameter value.  The proposal width was tuned
manually using preliminary runs with a procedure that estimates the conditional standard deviation in the posterior distribution \citep{Raftery1995}. This resulted in
different proposal widths between the two parametrisations.
We evaluated the performance of each MCMC algorithm by estimating the Effective
Sample Size (ESS) for each parameter using the LaplacesDemon package in R.  There is a wide range in the ESS values
across parameters.  This is because some parameters are largely independent of
the other parameters (leading to a high ESS).  However there are also strongly
correlated parameter pairs (leading to a low ESS).  When the model is
reparametrised the minimum ESS is around 14 times higher than it is with the
original parametrisation; the same is true of the average ESS (see Tab.~\ref{tab:ess}).

It is worth noting that the difference in the ESS is not the same for all parameter sets.  Indeed there are some scenarios where the ESS decreases for some of the parameters when the model is reparameterized (not shown).  It is also worth noting that the ESS in the reparametrised model is still only a small fraction of the number of
MCMC iterations, around $0.1-0.5\%$.  These
observations suggest that even when the model is reparamterised, the MCMC mixes
relatively slowly.  Depending on the model and computational resources available
this may or may not be a problem.  If it is problematic, as we have found to be
the case for the Neural Population Model considered later, then algorithms based
on Langevin or Hamiltonian dynamics may perform better.  If the mixing rate is
not problematic, MwG with reparameterization offers an easy to implement
alternative to Langevin and Hamiltonian Monte Carlo methods.  Since MwG does not
require costly derivative calculations, there may also be situations when the
overall computational efficiency is higher (as measured by ESS / CPU-time).

\subsection{Comparison between Kalman filter and Whittle likelihood on linear model}
\label{sec:kalman_v_whittle}

This section continues the analysis of the Whittle likelihood accuracy.  We compare the Whittle likelihood with the Kalman filter in a situation where the Kalman filter is exact up to discretization error.  The linear model that we use is derived by linearizing
the FHN equations around a stable fixed point.
As mentioned above, for $d=I_0=0$ in Eqs.~\eqref{eq:fhn1} and \eqref{eq:fhn2}, the system will always
have an equilibrium at $(V^*,w^*)=(0,0)$. As in the previous section, we assume here that $I(t)=0$ and
$P(t)$ is white noise. If we are only interested in $v(t)=V(t)-V^*$ we can eliminate $w(t)-w^*$ from the
linearized model to obtain a driven harmonic oscillator
\begin{equation} \label{eq:ho}
\left(\frac{d^2}{d t^2} + 2 \zeta\omega_0 \frac{d}{d t} + \omega_0^2\right) v(t) = P(t),
\end{equation}
\noindent where $2 \zeta\omega_0  \equiv a + c$ and $\omega_0^2 \equiv ac + b$. Note
that the conditions for obtaining a stable equilibrium point mentioned above then become
$2 \zeta\omega_0 >0$ and $\omega_0^2>0$. Without loss of generality, we hence can choose
$\omega_0>0$ for the natural frequency and then require $\zeta>0$ for the damping
to obtain stability.
Eq.~\eqref{eq:ho} is equivalent to what is sometimes referred to as the
continuous parameter AR(2) process \citep{Priestley1981}.  The spectral density
for the harmonic oscillator, driven by white noise, is a single peak if the oscillator
is sufficiently under-damped $\zeta<1/\sqrt{2}$.  The
location of the spectral peak is $\omega_1 =2\pi\nu_1=\omega_0 \sqrt{1
  - 2\zeta^2}$.

\begin{figure}[h]
\centering
\includegraphics[width=\textwidth]{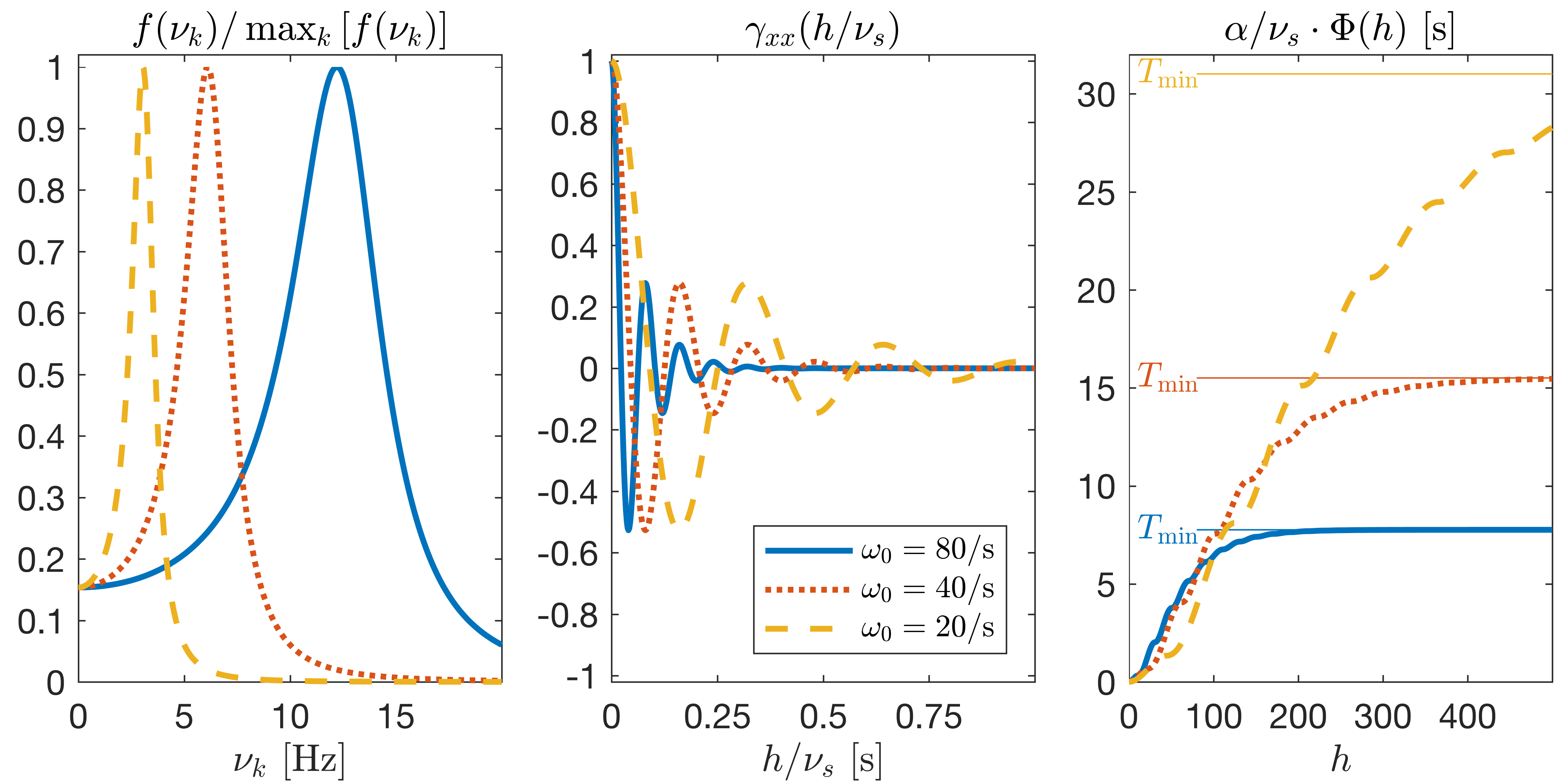}
\caption{Normed spectral density (\textit{left}), autocovariance (\textit{middle}), and accuracy heuristic (\textit{right}) for the harmonic oscillator with damping $\zeta = 0.2$ and three different $\omega_0$:
80/s (blue solid lines), 40/s (dotted red lines), and
20/s (golden dashed lines). Sampling frequency $\nu_s=500~\mathrm{Hz}$. The minimum time-series length can be calculated using Eq.~\eqref{eq:heuristic}:
$n_{\min}=\nu_s T_{\min}\equiv\alpha\phi=\alpha\lim_{h \rightarrow \infty}\Phi(h)$, where $\alpha \equiv ( 0.01 \cdot \max_k\left[ f(\nu_k)\right])^{-1}$ and $\Phi(h) \equiv 2\sum_{j=0}^h |j||\gamma_{xx}(j/f_s)|$. $T_{\min}$ values
for the different $\omega_0$ are shown as thin lines
in the \textit{right} panel in matching colours.}
\label{fig:autocorr}       
\end{figure}
\clearpage

In Sec.~\ref{sec:kalman_v_whittle}, we saw that the accuracy of of the
Whittle likelihood depends on two quantities: $\phi$ given by Eq.~\eqref{eq:supp_phi}
and $n$, the length of the time series.  Inspection of Eq.~\eqref{eq:supp_phi} reveals that
$\phi$ (and hence the error in the
Whittle likelihood) is larger when the autocovariance is high at long lags,
i.e., for strong correlations between time-points that are far away from each other
in the discrete-time index. Further analysis has been done for the AR(1) process,
\citep{Contreras-Cristan2006}, showing that a higher AR(1) coefficient (i.e.,
stronger correlation between time-points) lead to a larger error.  For processes
with a monotonically decaying autocovariance function it is clear that a slower
decay in the autocovariance will mean that a longer time-series is needed to
justify the use of the Whittle likelihood.

\begin{figure}[h]
\centering
\includegraphics[width=\textwidth]{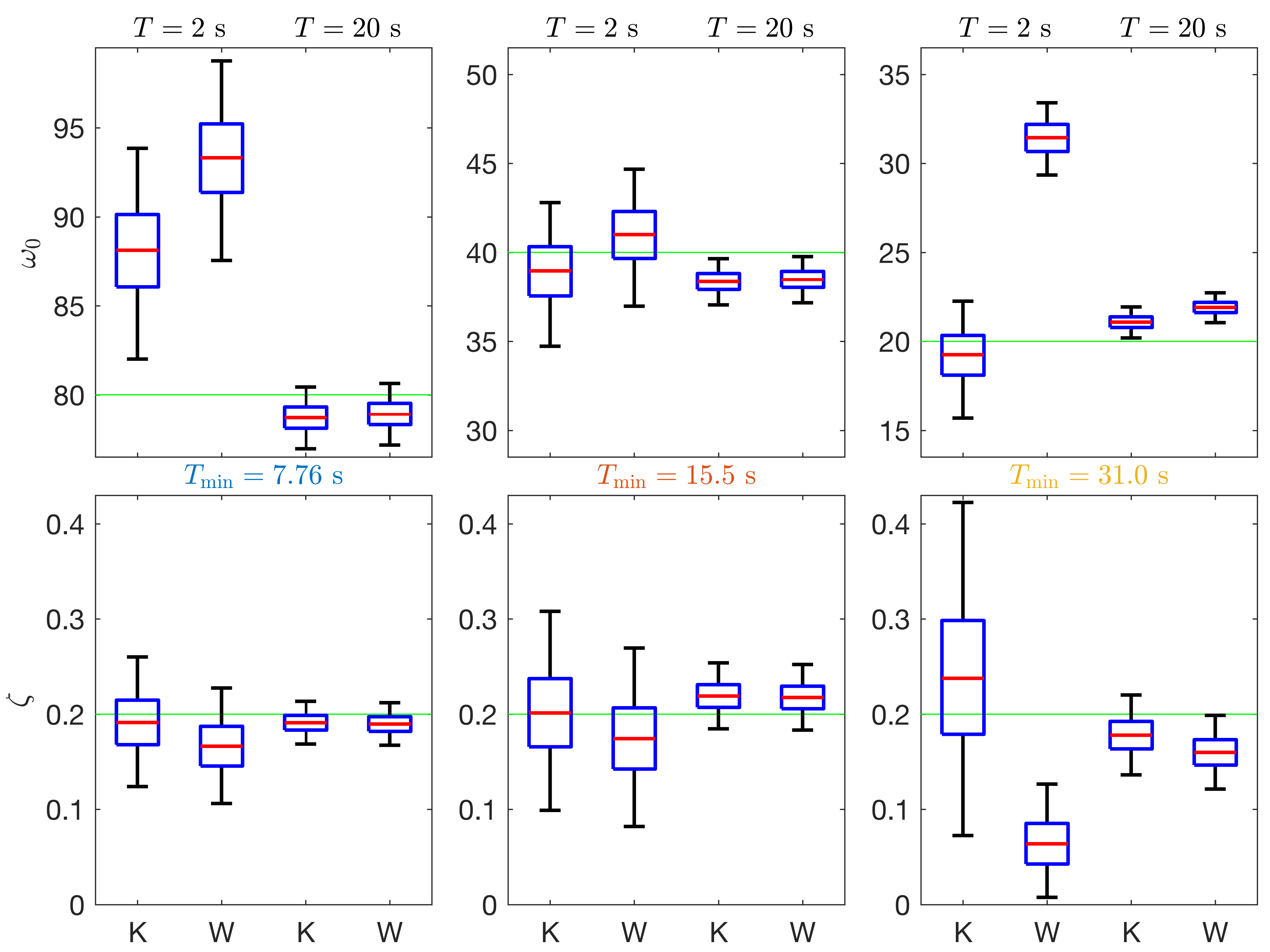}
\caption{Boxplots of the marginal posterior distributions for $\omega_0$ (\textit{top})
and $\zeta$ (\textit{bottom}) under a uniform prior.  The true
parameters values are $\zeta=0.2$ (all plots), and $\omega_0=80, 40,20/s$ (from \textit{left} to \textit{right}). The exact Kalman filter (K) and the
approximate Whittle likelihood (W) analyses are in agreement
only if $T\geq T_{\min}$, as predicted by the
heuristic, Eq.~\eqref{eq:heuristic}, cf.\ Fig.~\ref{fig:autocorr}
\textit{right}.}
\label{fig:2}       
\end{figure}

In the case of oscillatory processes, such as the linearized stochastic FHN equations and
the neural population model discussed in Sec.~\ref{sec:npm}, the autocovariance function is oscillatory and
decays to zero.  There are then two distinct factors which affect the magnitude of $\phi$.
First is the frequency of oscillations, or equivalently the location of resonance peak(s)
in the spectral density.  If the systems oscillates at a lower frequency this
results in an autocovariance function with a longer period.  Second is the
regularity of the oscillations, or equivalently the width of the resonance
peak(s) in the spectral density.
If the frequency of oscillations is highly consistent over time, as opposed to
being more spread across frequencies, this results in an autocovariance
function that decays more slowly.

Further quantification of the error can be done by comparing likelihoods from
the Kalman filter (which is exact up to discretization error) with the Whittle
likelihood.  For our purposes, the Whittle likelihood is accurate when the
posterior distribution under the Whittle likelihood is similar to the posterior
distribution under the Kalman filter.
By numerical experimentation, we have found that
the Kalman filter and Whittle likelihood
posteriors are similar for the linearized FHN equations when
\begin{equation}
\frac{\phi}{n} < 0.01 \max_k\left[ f(\nu_k)\right].
\end{equation}
This is illustrated in Fig.~\ref{fig:2}. Since it is quicker to compute the quantity $\phi / n$ than to do a full
comparison between posterior distributions, we recommend using Eq.~\eqref{eq:heuristic} as a heuristic
rule to decide whether the Whittle likelihood is
a sufficiently accurate approximation to the true likelihood.
For the example in the next section, estimating the posterior distribution using the Kalman filter is intractable and we rely on the heuristic in Eq.~\eqref{eq:heuristic} to give an indication of whether the Whittle likelihood is accurate.

\begin{figure}[ht]
\centering
\includegraphics[width=\textwidth]{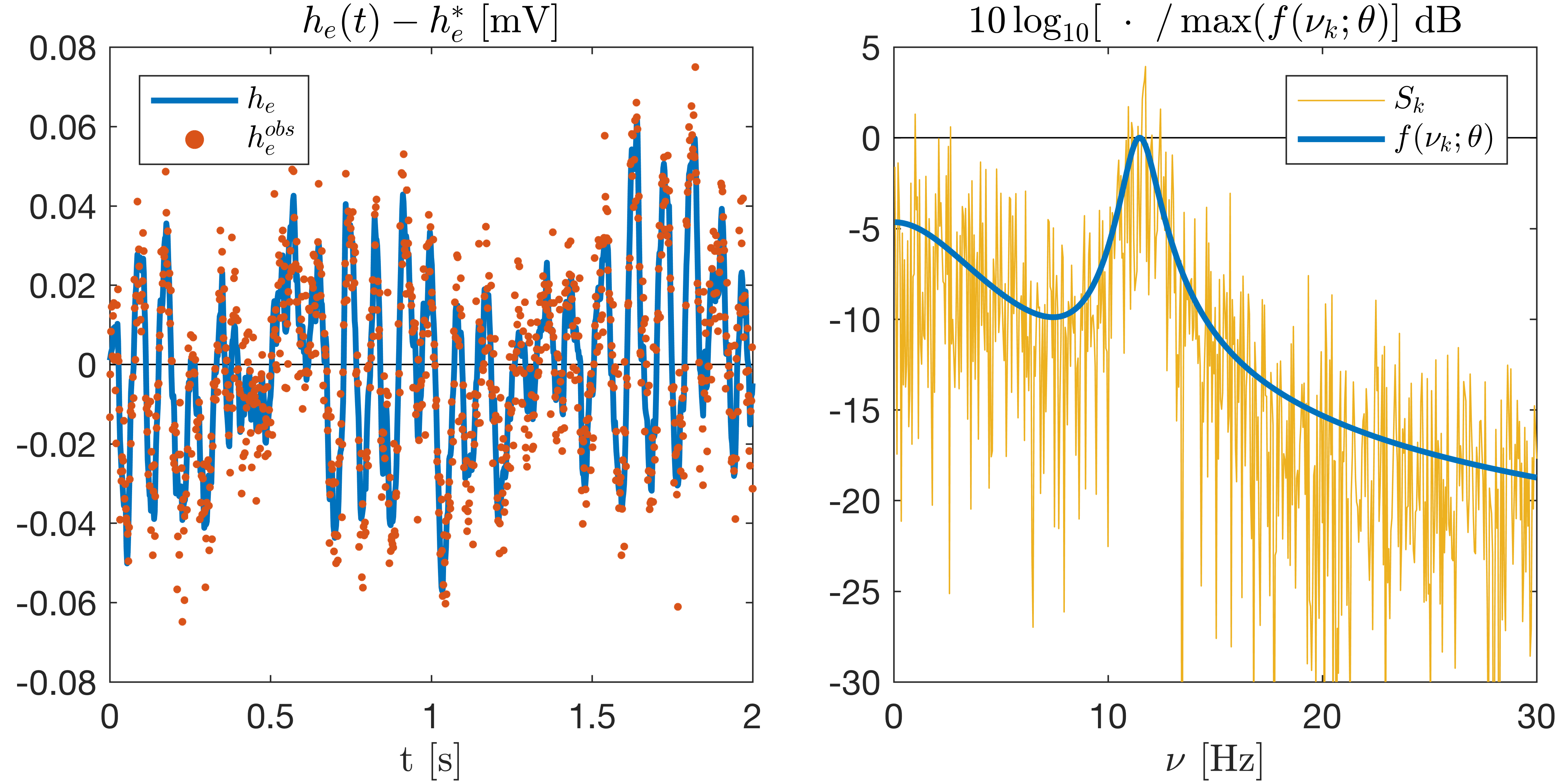}
\captionof{figure}{\textit{Left:} First 2~s of $T=20$~s of pseudo-data generated with parameters from Tab.~\ref{tab:npm} as solid blue line, with noisy observations thereof as red dots. The (equilibrium) mean has been subtracted.  \textit{Right:} Decibel plot of spectral density of $T=20$~s observed pseudo-data (thin golden line) and predicted spectral density of linearized NPM model (thick blue line).}
\label{fig:spec}
\end{figure}
\clearpage

\section{Analysis using Neural Population Model}

\subsection{Model description}
\label{sec:modeldef}

The model in this section describes the mass action of neural tissue, rather than the activity of individual neurons. This class of models carries a variety of labels in the literature -- usually called a ``mass model'' if discrete units of tissue are considered, but a ``field model'' if the tissue is modeled as a continuum. The model used here can represent either case, depending on how the activity propagation Eq.~\eqref{eq:prop} is specified, and we will use the generic label neural population model (NPM).
We are interested in the neuroimaging problem of inferring NPM model
parameters from EEG data.  The characteristics of EEG datasets vary
depending on the experimental setup and experimental subjects being
studied.  Event Related Potentials (ERPs) and epileptic seizure
dynamics typically feature non-stationary and/or nonlinear behaviour
\citep{David2005, Andrzejak2001}.  In contrast, adult resting-state
data can be considered as weakly stationary \citep{Kreuzer2014} with a
power spectrum that is a superposition of an inverse frequency
(typically between $1/f$ and $1/f^2$) curve with a resonance peak in
the so-called alpha range (8-13~Hz).
Here we restrict ourselves to inferring the parameters
when the true parameter values are static as opposed to time-varying.

Our NPM is known as the Liley model \citep{Liley2002, Bojak2005}, and it distinguishes two neural populations by their function: one excitatory, the other inhibitory.
That distinction is made according to the effect neurons have on other neurons they are connected to. If the activity of the source neurons increases the activity of the target neurons, then they are excitatory, if it decreases it, inhibitory.
The model can be split into three parts: dynamics of the
mean soma membrane potentials ($h_e$ and $h_i$), local reaction to synaptic inputs ($I_{ee}$, $I_{ei}$, $I_{ie}$, and $I_{ii}$), and
long-range propagation of activity ($\Phi_{ee}$ and $\Phi_{ei}$).
In modeling the EEG with this NPM, one typically assumes that the EEG observations are linearly proportional to $h_e$ with some added observational noise.

The equations for the membrane potential dynamics are as follows,
\begin{equation}
\label{eq:lb1}
\begin{aligned}
\tau_e \frac{\partial}{\partial t} h_e(\mathbf{x},t) &= h_e^r - h_e(\mathbf{x},t) + \psi_{ee}\left[h_e(\mathbf{x},t)\right] I_{ee}(\mathbf{x},t) + \psi_{ie}\left[h_e(\mathbf{x},t)\right] I_{ie}(\mathbf{x},t),\\
\tau_i \frac{\partial}{\partial t} h_i(\mathbf{x},t) &= h_i^r - h_i(\mathbf{x},t) + \psi_{ei}\left[h_i(\mathbf{x},t)\right] I_{ei}(\mathbf{x},t) + \psi_{ii}\left[h_i(\mathbf{x},t)\right] I_{ii}(\mathbf{x},t),
\end{aligned}
\end{equation}

\noindent where

\begin{equation}
\psi_{kl}[h_l(\mathbf{x},t)] \equiv \frac{h_{kl}^{eq} - h_{l}(\mathbf{x},t)}{|
h_{kl}^{eq} - h_{l}^r |}.
\end{equation}

\noindent The subscript $e$ stands for excitatory, $i$ for inhibitory, and $k$ and $l$ can take on either of these values. A single subscript denotes a quantity relating to a single neural population; whereas a double subscript denotes a quantity relating to the interaction between two populations, i.e., subscript $kl$ indicates $k$ acting on $l$.
In the absence of synaptic inputs $I_{kl}$ the membrane potentials $h_l(\mathbf{x},t)$ will decay to rest values $h_l^r$ with characteristic times $\tau_l$.
Synaptic inputs $I_{kl}\geq 0$ are weighted by the functions $\psi_{kl}$. Since for
biological parameter ranges $h_{el}^{eq}>h_{l}^r$ but $h_{il}^{eq}<h_{l}^r$,
at rest excitatory inputs $I_{el}$ are weighted by $+1$ but inhibitory ones
$I_{il}$ by $-1$. Note that possibility of having zero weight or even
a reversal of the sign, depending on the current state of the membrane potential,
is not a mathematical artifact but reflects properties of voltage-gated ion channels.

The equations for local synaptic impact are

\begin{equation}
 \label{eq:lb2}
\begin{aligned}
\left( \frac{1}{\gamma_{el}}\frac{\partial }{\partial t} + 1 \right)^2 I_{el}(\mathbf{x},t) &=
q_{el} \left\{N^{\beta}_{el} S_e\left[h_e(\mathbf{x},t)\right] + \Phi_{el}(\mathbf{x},t) + p_{el}(\mathbf{x},t) \right\} \\
\left( \frac{1}{\gamma_{ik}}\frac{\partial }{\partial t} +  1\right)^2 I_{ik}(\mathbf{x},t) &=
q_{ik}  \left\{N^{\beta}_{ik} S_i\left[h_i(\mathbf{x},t)\right]\right\}
\end{aligned}
\end{equation}

\noindent where

\begin{gather}
p_{ee}(\mathbf{x},t)=\bar{p}_{ee}+p(\mathbf{x},t),\qquad
p_{ei}(\mathbf{x},t)=\bar{p}_{ei}~(\mathrm{const.})\\
S_k\left[h_k(\mathbf{x},t)\right] = S_k^{max}/\left\{1 + \exp\left[
\frac{\bar{\mu}_k - h_k(\mathbf{x},t)}{\hat{\sigma}_k/\sqrt 2} \right]\right\}.
\end{gather}

\noindent If the curly braces in Eqs.~\eqref{eq:lb2} are
replaced by a single Dirac delta distribution $\delta(t)$ representing one
input spike, then one obtains as postsynaptic response canonical ``alpha functions''
$\gamma^2t \exp(-\gamma t)\Theta(t)$ weighted by $q$, where the indices
have been left out for clarity and $\Theta(t)$ is the Heaviside step distribution.
The terms in the curly braces hence represent counts for sources of such
input spikes: local cortical firing $S_k$, firing propagated long-range
in cortex $\Phi_{el}$ and extra-cortical input $p_{el}$ in particular from the
thalamus.

The equations for the long-range propagation across cortex are
\begin{equation}
\label{eq:prop}
\left[ \left(\frac{1}{v\Lambda}\frac{\partial}{\partial t} + 1\right)^2 - \frac{3}{2\Lambda^2}
\nabla^2 \right] \Phi_{ek}(\mathbf{x},t) = N^{\alpha}_{ek} S_e\left[h_e(\mathbf{x},t)\right]
\end{equation}

\noindent These equations correspond approximately to activity propagating isotropically outwards from firing sources $S_e\left[h_e(\mathbf{x},t)\right]$ with conduction velocity $v$, while decaying exponentially with a characteristic length scale $\Lambda$. Note that this would correspond to a ``field model'' in the
terminology introduced at the beginning of this section. However, for
simplicity we assume in what follows that the the system is spatially homogeneous, i.e., for the input $p_{ee}(\mathbf{x},t) \equiv p_{ee}(t)$ and likewise
for all state variables. This describes coherent bulk oscillations of the brain. Then $\nabla^2\Phi_{ek}(\mathbf{x},t) \equiv 0$, and Eq.~\eqref{eq:prop} becomes
a damped harmonic oscillator ODE without spatial dependence.
Consequently, one can also interpret this
as a ``mass model'' for a single unit of tissue. Alternatively, one can obtain a ``mass model'' by simply dropping Eq.~\eqref{eq:prop} entirely from the Liley model. However, the homogeneous ``mass model'' typically produces activity more similar
to the ``field model'' for the same parameter values. The methodology developed in this paper can be applied to spatial models as well.  See \cite{Bojak2005} for details on how to evaluate the spectral density for spatially dependent solutions.

\begin{table}[h]
\begin{center}
\begin{tabular}{|c|r||c|r||c|r||c|r|}
\hline
\multicolumn{2}{|c||}{NPM knowns} & \multicolumn{2}{|c||}{NPM knowns} & \multicolumn{2}{|c||}{NPM unknowns}        & \multicolumn{2}{|c|}{auxiliary} \\
\hline\hline
$\tau_e$          &  105.5 ms & $S^{max}_e$       &  355.2/s    & $\gamma_{ee}$  & 841.2/s              & \multicolumn{2}{|l|}{sampling frequency} \\
$\tau_i$          &  149.0 ms & $S^{max}_i$       &  424.8/s    & $\gamma_{ei}$  & 859.7/s              & $f_s$          &  500 Hz \\
$h^r_e$           &  -71.9 mV & $\bar{\mu}_e$     &  -52.3 mV   & $\gamma_{ie}$  & 451.7/s              & \multicolumn{2}{|l|}{} \\
$h^r_i$           &  -76.0 mV & $\bar{\mu}_i$     &  -48.6 mV   & $\gamma_{ie}$  & 451.7/s              &  \multicolumn{2}{|l|}{recording duration} \\
$h^{eq}_{ee}$     &   -9.0 mV & $\hat{\sigma}_e$  &    4.9 mV   & $q_{ee}$       &  $0.9484~\mu\Omega\mathrm{C}$ & $T$            &   20 s \\
$h^{eq}_{ei}$     &  -10.3 mV & $\hat{\sigma}_i$  &    3.7 mV   & $q_{ei}$       &  $5.835~\mu\Omega\mathrm{C}$  & \multicolumn{2}{|l|}{} \\
$h^{eq}_{ie}$     &  -87.5 mV & $v$               &  780.7 cm/s & $q_{ie}$       & $11.99~\mu\Omega\mathrm{C}$   & \multicolumn{2}{|l|}{std. dev. observation noise} \\
$h^{eq}_{ii}$     &  -82.3 mV & $\Lambda$         &    0.345/cm & $q_{ii}$       & $11.30~\mu\Omega\mathrm{C}$   & $\sigma_{obs}$ &    0.01 mV \\
$N^{\beta}_{ee}$  & 2194      & $N^{\alpha}_{ee}$ & 3668        & $\bar{p}_{ee}$ & 6025/s                & \multicolumn{2}{|l|}{}  \\
$N^{\beta}_{ei}$  & 4752      & $N^{\alpha}_{ei}$ & 1033        & $\bar{p}_{ei}$ & 1116/s                & \multicolumn{2}{|l|}{std. dev. noise input} \\
$N^{\beta}_{ie}$  &  700      & & & & &  $\sigma_p$     &  100.0/s  \\
$N^{\beta}_{ii}$  &  516      & & & & & \multicolumn{2}{|l|}{} \\
\hline
\end{tabular}
\end{center}
\captionof{table}{List of NPM parameter values used to generate pseudo-data, taken from column~3 of Table~5 in \cite{Bojak2005}, as well as other parameters used in the simulation run. In the inference problem parameters labeled ``NPM knowns'' are assumed as given, i.e., only the ``NPM unknowns'' are being inferred from the pseudo-data.}
\label{tab:npm}
\end{table}

For numerical calculations the noise input was treated with the Euler-Maruyama method, i.e., using independently generated Gaussian white noise in every time step
\begin{equation}
\ldots+h_t\cdot \bar{p}_{ee}+\frac{\sigma_p}{\sqrt{10^{-4}/s}}\sqrt{h_t}\cdot{\cal N}(0,1),
\end{equation}
where we have normed to the standard numerical time step $10^{-4}~\mathrm{s}$ for convenience in interpreting the noise standard deviation $\sigma_p$. Observations are made of the mean excitatory soma membrane potential $h_e$, and to simulate experimental measurement error we add independent Gaussian white noise with standard deviation $\sigma_{obs}$
\begin{equation}
h_e^{obs}(t_k)=h_e(t_k)+\sigma_{obs}\cdot{\cal N}(0,1).
\end{equation}
Here $t_k=k\cdot\Delta t$ with $k=0,\ldots,N-1$ indicates the
discrete observations of the system. An illustration of the pseudo-data and predictions from this NPM is given in Fig.~\ref{fig:spec}.

\subsection{Model reparameterization}
\label{sec:npm_reparam}
For the NPM defined in Sec.~\ref{sec:modeldef}, the equations of the form
\eqref{eq:ss2as} can be obtained by re-arranging the steady-state equations for
short-range and long-range connectivity,

\begin{align}
I^*_{ee} &= q_{ee} \left[N^{\beta}_{ee} S_e(h_e^*) + \Phi^*_{ee} + \bar{p}_{ee} \right] ,\label{eq:npmss_a}\\
I^*_{ei} &= q_{ei} \left[N^{\beta}_{ei} S_e(h_e^*) + \Phi^*_{ei} + \bar{p}_{ei} \right],   \\
I^*_{ik} &= q_{ik} \left[N^{\beta}_{ik} S_i(h_i) \right],
\\
\Phi^*_{ee} &= N^{\alpha}_{ee} S_e(h_e^*),  \\
\Phi^*_{ei} &= N^{\alpha}_{ei} S_e(h_e^*).\label{eq:npmss_c}
\end{align}

\noindent The remaining steady-state equations are,

\begin{align}
0 &=  h_e^r - h^*_e + \psi_{ee}(h^*_e)
I^*_{ee} + \psi_{ie}(h^*_e) I^*_{ie}, \label{eq:lbss1a} \\
0 &= h_i^r - h^*_i + \psi_{ei}(h^*_i)
I^*_{ei} + \psi_{ii}(h^*_i) I^*_{ii}.\label{eq:lbss1b}
\end{align}

\noindent Equations $\eqref{eq:npmss_a}-\eqref{eq:npmss_c}$ can be used to eliminate
$I^*_{ee}$, $I^*_{ei}$, $I^*_{ie}$, $I^*_{ii}$ from \eqref{eq:lbss1a}-\eqref{eq:lbss1b}.  The resulting equations are of the form \eqref{eq:ss2bs} with $\mathbf{x}^*_2 = (h_e^*, h_i^*)$, and $\theta_d =
(\bar{p}_{ee}, \bar{p}_{ei})$. Note that the solutions for $\theta_d$ are easily obtained in closed form.

\subsection{Parameter values for prior distribution}
\label{sec:npm_prior}

\begin{table}[ht]
  \centering
\begin{tabular}{| c | c | c | c |}
  \hline
$\gamma_{ee}$ & 500/s & $q_{ee}$ & $5.437~\mu\Omega\mathrm{C}$  \\
$\gamma_{ei}$ & 500/s & $q_{ei}$ & $5.437~\mu\Omega\mathrm{C}$ \\
$\gamma_{ie}$ & 250/s & $q_{ie}$ & $10.87~\mu\Omega\mathrm{C}$ \\
$\gamma_{ii}$ & 250/s & $q_{ii}$ & $10.87~\mu\Omega\mathrm{C}$ \\
$\bar{p}_{ee}$ & 5000/s & $\bar{p}_{ei}$ & 5000/s \\
  \hline
\end{tabular}
  \captionof{table}{Modes of prior distribution components. }
  \label{tab:prior}
\end{table}

The prior distribution component modes for $\gamma_{ee}$, $\gamma_{ei}$, $\gamma_{ie}$, $\gamma_{ii}$, $q_{ee}$, $q_{ei}$, $q_{ie}$, and $q_{ii}$ are shown in Tab.~\ref{tab:prior}.  In addition we treated the variance of the input noise as an unknown parameter (denoted by $\sigma_p^2$), also with a log-normal prior distribution.  The mode of the prior distribution for this noise variance was $10^4/s^2$.   For most of the parameters we chose the standard deviation of the distribution on the log scale to be 2.  For the mean inputs, $\bar{p}_{ee}$ and $p_{ei}$, the log standard deviation was set to $0.5$.  Physiologically, the time-varying input should be positive, so putting a relatively tight prior distribution on the mean ensures that this will be the case most of the time.
The prior parameters we have chosen put a lot of prior mass on values above the upper limits in the physiological range.  Within the physiological ranges stated in \cite{Bojak2005}, the prior density is relatively uniform (it is always at least half of the maximum prior density).  Ideally we may prefer prior distributions that take better account of the physiological ranges.  Instead we have used here a pragmatic solution that errs on the side of non-informativeness.

\bibliography{library_main,library_supp}

\end{document}